\documentclass{aa}
\usepackage{txfonts}
\usepackage{natbib}
\usepackage{graphicx}
\usepackage{aalongtable}
\bibpunct{(}{)}{;}{a}{}{,}
\begin{document}

\title{The lack of close binaries among hot horizontal branch
stars in \object{NGC6752}
\thanks{Based on observations with the ESO Very Large Telescope
at Paranal Observatory, Chile (proposal ID 69.D-0682)}
}

\author{
C. Moni Bidin \inst{1}\fnmsep\inst{2}
\and
S. Moehler \inst{3}
\and
G. Piotto \inst{1}
\and
A. Recio-Blanco \inst{1}\fnmsep\inst{4}
\and
Y. Momany \inst{1}
\and
R.A. M\'{e}ndez \inst{2}
}

\institute{
Dipartimento di Astronomia, Universit\`{a} di Padova,
Vicolo dell'osservatorio 2, 35122 Padova, Italy
\and
Departamento de Astronom\'{i}a, Universidad de Chile,
Casilla 36-D, Santiago, Chile
\and
Institut f\"{u}r Theoretische Physik und Astrophysik,
Christan-Albrechts-Universit\"at zu Kiel, 24098 Kiel, Germany
\and
Observatoire de la C\^{o}te d'Azur,
Dpt. Cassiop\'ee, CNRS UMR 6202,
B.P. 4229, 06304 Nice, Cedex 04, France
}
\date{Received / Accepted }


\abstract
{}
{We present the results of a spectroscopic search for
close binaries among horizontal branch (HB)
stars in \object{NGC6752}.}
{We used the ESO VLT-FORS2 instrument to obtain medium resolution
(R=4100) spectra of
51 hot HB stars with $8\,000\ $ K $ \leq \mathrm{T_{eff}} \leq 32\,000\ $ K
during four consecutive nights.
Eighteen of our targets are extreme horizontal branch (EHB) stars
with $\mathrm{T_{eff}}\ge 22\,000\ $K.
Radial velocity variations were measured with
cross-correlation techniques and we carefully evaluated the
statistical and systematic errors associated with them.}
{No close binary system has been detected
among our 51 targets.
The data corrected for instrumental effects indicate that
the radial velocity variations
are always below $\approx$ 15 km s$^{-1}$ (3$\sigma$ level).
From a statistical analysis
of our results, we
conclude that (at 95\% confidence level) the fraction
of binaries with a $\sim0.5\ \mathrm{M_{\sun}}$ companion
among EHB stars in \object{NGC6752} is smaller than 20\%.}
{This empirical evidence
sharply contrasts with what has been found for
hot subdwarfs in the  field, and opens new questions about the formation
of EHB stars in globular clusters (and possibly in the field as well).}

\authorrunning{Moni Bidin et al.}
\maketitle


\section{Introduction}
\label{capintro}

Although stellar evolution theory has successfully identified
horizontal branch (HB) stars as post-helium flash stars of low
( typically $0.7-0.9\ \mathrm{M_{\sun}}$ in globular clusters) initial
mass \citep{HoyESchwarz55,Faulk66}, we still lack a comprehensive
understanding of their nature. The large morphological differences
\citep[see for example][]{Piotto02}
among HBs of Galactic globular clusters (GCs) are surely one of the
most puzzling problems of stellar evolution. It is well known that
cluster metallicity affects the HB morphology \citep{SandEWalle60},
but there are still many morphological differences among the HBs
of GCs
with the same metallicity \citep{SandEWild67,VDBergh67},
  suggesting the presence of one or more second parameters. Many
"second
parameter" candidates have been proposed, but none has provided an
overall explanation for all the available observations, and it is
likely that the so-called ``second parameter problem'' is the
consequence of a combination of parameters \citep{Fusi93}.

In more recent years, more and more GCs have been found to show an HB
blue tail extending toward increasing effective temperatures and
fainter visual magnitudes \citep{Sosin97,Rich97,Ferraro98,Piotto99}. This
extreme horizontal branch (EHB) population with $\mathrm{T_{eff}} \geq
20\,000$ K is a challenging test of the theoretical models of the late
evolutionary stages of low-mass stars, and their origin is still not
understood (see below). The EHBs might also have important implications for
extragalactic astronomy, since EHB stars have been
identified as possibly being responsible for the UV
upturn in elliptical galaxies.
This UV upturn corresponds to the increase in
 flux with decreasing wavelength below
2500~\AA\ discovered in the bulge of M31 \citep{Code69}, and then found
in almost all bright elliptical galaxies and many spiral bulges.  It
has been proposed that the EHB stars are the source of this
increase in UV flux \citep[see for
example][]{GreERenz90,GreERenz99,Brown00}.  This idea has been
reinforced by the discovery of EHB stars in metal-rich GCs like \object{NGC6388}
and \object{NGC6441} \citep{Rich97} and in the Galactic bulge \citep{Buss05}.
As EHB stars evolve with time, the upturn is
expected to change during the evolution of the galaxies, and to be
almost absent for younger ellipticals
\citep{GreERenz90,Tantalo96}. A knowledge of the formation mechanisms
and evolution of EHB stars in different environments is therefore
 urgently needed in
order to model the behavior of the UV upturn with age (redshift). The
UV upturn could then also potentially be used as
an age discriminator for elliptical
galaxies.

While it is agreed that EHB stars are He-burning stars that have
suffered heavy mass loss during their evolution
\citep{IbenERood70,Faulk72,DCruz96}, keeping only a thin external
envelope with a mass around $0.02\ \mathrm{M_{\sun}}$ or lower,
their actual formation mechanism remains unclear.  Some mechanism able
to enhance the mass loss must be responsible for their formation. As
first proposed by \citet{WilEBow84}, many authors explored
mass-loss mechanisms of the HB itself.  \citet{Yong00}
successfully reproduce the morphology of blue HBs in metal-rich GCs
with a constant mass-loss rate during HB evolution, but provide
no explanation
for the required rate. \citet{VinkECass02} show that
enhanced radiation-driven winds cannot account for the high mass-loss
rates required, concluding that the most plausible mechanisms are unable to
form EHB stars by mass loss on the HB. On the other hand, many authors
propose heavy mass-loss rates during the previous red giant
branch (RGB) phase \citep[see for example][]{Soker01}. \citet{Iben90}
propose an alternative scenario, in which the EHB stars are formed
from mergers of helium white-dwarf binary systems. \citet{Fusi93}
indicate that the GC density can enhance the presence of EHB stars,
hinting that dynamical interactions may play an important role in
their formation.

There is still a lack of models that can explain such heavy mass
loss in the evolution of a single low-mass star.  The binarity of EHB
stars, as proposed by many authors \citep{Mengel76,Heber02}, might
provide an explanation for their formation, since the dynamical
interaction with a close compact companion can enhance the mass loss
through a number of different binary-evolution channels
\citep[][ see below]{Han04},
particularly during the RGB phase.  The presence of a binary
population in GCs is now well established \citep{Hut92,Bail93}, and
binaries can get closer and closer as a consequence of the
dynamical interactions inside GCs \citep{Heggie75}.  The idea that EHB
stars are components of close binary systems has been strengthened by
observations of field EHB stars, also known as subdwarf B-type stars
(sdBs). Binaries have been found to be very common among
them. \citet{Maxted01} conclude from their observations that $69 \pm 9
\%$ of sdB stars are binaries.
\citet{Moral03} recently measured the orbital periods P
and the semiamplitudes of radial velocity variation K of 22 new binary
sdBs, increasing the number of sdBs for which these quantities are
known to 38. Thirty of them have periods below 3 days
 (22 with P $\leq 1$ day),
showing that they are almost all very close systems, with K usually
exceeding 50 km s$^{-1}$ (only 7 exceptions) and easily greater than
100 km s$^{-1}$.
A more recent search for binaries among field sdB stars \citep{Napiw04} led
to a much lower (42\%) close binary fraction than expected from
previous results. \citet{Napiw04} point out that their sample
contained a much higher percentage of faint, i.e. distant, stars
  than the one of \citet{Maxted01}. The Napiwotzki et al. sample
  may therefore extend to greater distances from the Galactic plane and
  thus be contaminated
by thick-disk or halo members. This could imply that a relation
  of binary frequency with
metallicity and/or age is present.

\citet{Han03} find from binary
population synthesis techniques that 76--89\% of the sdBs should
be close binaries.
\citet{Han04} analyzed in detail the
main binary evolution channels that can lead to the sdB formation.
Comparing their models with the available
empirical data, they find that a very efficient
common-envelope (CE)
channel
can fit the observed distribution of close-binary sdB
periods, though it is not possible to fit the observed cumulative
luminosity functions, as shown by \citet{Lisker05}, showing that there
are still some problems with the binary scenario.
In the CE channel, the progenitor of the sdB star is a giant that
fills its Roche lobe near the tip of the red giant branch and
experiences mass transfer during which the core of the giant and the
companion (a normal main-sequence star or a white dwarf) spiral
towards each other inside a common envelope formed out of the giant
envelope. Once enough orbital energy has been released, the envelope is
ejected and the system becomes a short-period binary.

There are alternative mechanisms suggested for the formation of
the EHB stars. Recent observational results on \object{$\omega$
Centauri} \citep{Bedin04,Piotto05}, \object{M3}-\object{M13}
\citep{CalEDant05}, and \object{NGC2808}
\citep{dantona05} have given new impulse to the idea that EHB
stars may result from a
second generation of stars enriched in helium by pollution from
intermediate-mass AGB star ejecta \citep{Dantona02,Lee05}.

In the attempt to shed some light on the origin of EHB stars in
  globular clusters, we want to test in this
paper the possibility that they are stars that
have experienced greatly enhanced mass loss during the evolution in a
close binary system.
We started our spectroscopic search for close binaries among the EHB
stars of \object{NGC6752}. This GC is an ideal target as it is
dynamically evolved, shows an extended and well-populated blue HB (as
shown in Fig. \ref{targethr}), a large population of X-ray sources
in the core that could be cataclysmic variables \citep{Pool02}, and it
is supposed to have a large main-sequence binary population
\citep{RubeEBail97}.

In a forthcoming paper, we will present and discuss the main
atmospheric parameters ($\mathrm{T_{eff}}$, log$g$, $\log
{\frac{n_{\rm He}}{n_{\rm H}}}$, mass) derived from a complementary
set of spectra obtained during the same observational run. In the
following, we concentrate on the radial velocity and radial
velocity variation measurements, and on their implication for the
close binary scenario in the formation of EHB stars.


\section{Observation and data reduction}
\label{capdata}

\begin{figure}
\begin{center}
\resizebox{\hsize}{!}{\includegraphics{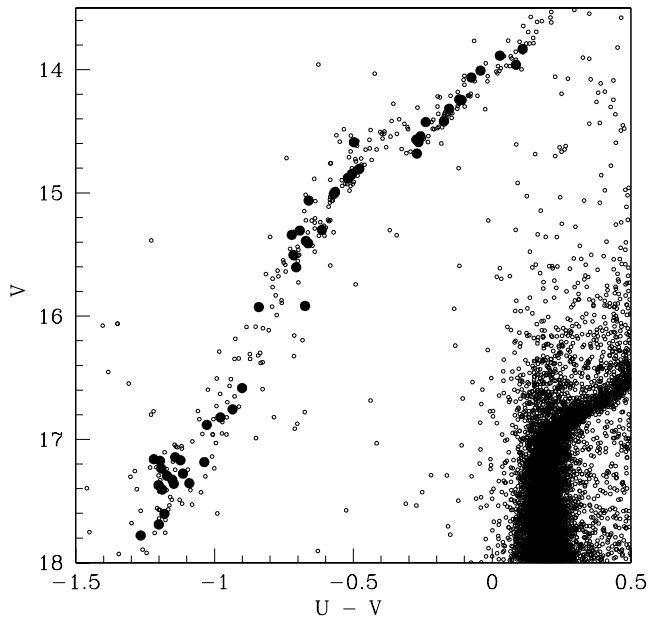}}
\caption{Positions of target stars (solid dots) in the
color-magnitude diagram of \object{NGC6752}. Data from \citet{Moman02}.}
\label{targethr}
\end{center}
\end{figure}

The spectra were acquired during four nights of observation, from June
11 to June 14, 2002, at the VLT-UT4 telescope equipped with the FORS2
spectrograph in MXU mode.  Fifty-one HB stars were selected from the
photometric catalog of \object{NGC6752} by \citet{Moman02}.  The stars
are well distributed along the HB from the cool edge ($\mathrm{T_{eff}}
\approx 8\,000~K$) to the EHB ($\mathrm{T_{eff}} \approx 30\,000~K$),
and were divided into 3 fields for multi-object spectroscopy. The
position of our targets on the HB are indicated in Fig.
\ref{targethr} and their radial distribution is shown in
  Fig.~2.

\begin{table*}[h!]
\begin{center}
\caption{List of the observed stars with their physical parameters.
IDs, coordinates (RA and DEC), magnitudes V and color ($U-V$) are
from \citet{Moman02}. The absolute heliocentric radial velocities ($V_{rad}$)
were measured in the
present work, the effective temperatures were obtained from the $U-V$ color
of the targets, with the relation ($U-V$) vs. T$_{\mathrm{eff}}$ shown in
Fig.\ref{teffM00} and described in the text.
The estimated error for temperatures is $\sigma_{\mathrm{T}}$= 5\%}
\label{targetlist}
\begin{tabular}{ c r r c c c r r r }
\hline \hline
field & slit & ID & RA (J2000) & DEC (J2000) & V & $V_\mathrm{rad}$ & $\mathrm{T_{eff}}$ & ($U-V$) \\
 & & & hh:mm:ss & $^{\circ}$: ' : '' & & km s$^{-1}$ & K & \\
\hline
A & 1 & 14770 & 19:11:32.563 & $-$59:59:37.63 & 17.247 & $-35\pm10$ & 30300 & $-$1.192 \\
A & 2 & 11634 & 19:11:16.084 & $-$60:00:27.33 & 14.431 & $-43\pm10$ & 10000 & $-$0.174 \\
A & 3 & 14944 & 19:11:22.485 & $-$59:59:35.51 & 15.401 & $-44\pm10$ & 15600 & $-$0.671 \\
A & 4 & 15026 & 19:11:12.319 & $-$59:59:34.51 & 14.018 & $-36\pm9$ & 9100 & $-$0.042 \\
A & 5 & 16551 & 19:11:14.253 & $-$59:59:13.90 & 15.420 & $-44\pm10$ & 15500 & $-$0.663 \\
A & 6 & 15395 & 19:11:05.940 & $-$59:59:29.51 & 17.337 & $-40\pm8$ & 29000 & $-$1.153 \\
A & 7 & 20919 & 19:11:10.496 & $-$59:58:15.25 & 13.844 & $-34\pm7$ & 8100 & $+$0.110 \\
A & 8 & 18782 & 19:11:16.428 & $-$59:57:45.90 & 14.601 & $-35\pm10$ & 13100 & $-$0.498 \\
A & 9 & 17941 & 19:11:04.827 & $-$59:58:02.02 & 15.928 & $-39\pm8$ & 16700 & $-$0.674 \\
A & 10 & 20302 & 19:11:16.912 & $-$59:57:13.20 & 16.594 & $-36\pm7$ & 20800 & $-$0.901 \\
A & 11 & 26756 & 19:11:25.759 & $-$59:56:17.07 & 14.599 & $-21\pm9$ & 10800 & $-$0.265 \\
A & 12 & 27181 & 19:11:24.177 & $-$59:56:04.10 & 15.073 & $-34\pm10$ & 15400 & $-$0.661 \\
A & 13 & 24849 & 19:10:59.399 & $-$59:57:05.99 & 14.578 & $-49\pm9$ & 10800 & $-$0.273 \\
A & 14 & 27604 & 19:11:17.021 & $-$59:55:50.56 & 15.937 & $-36\pm9$ & 19200 & $-$0.841 \\
A & 15 & 28231 & 19:11:18.560 & $-$59:55:27.51 & 17.367 & $-29\pm6$ & 26900 & $-$1.091 \\
A & 16 & 26760 & 19:10:59.788 & $-$59:56:17.51 & 15.515 & $-37\pm7$ & 16400 & $-$0.716 \\
A & 17 & 28554 & 19:11:07.939 & $-$59:55:14.16 & 17.289 & $-19\pm8$ & 27600 & $-$1.114 \\
A & 18 & 28693 & 19:11:08.108 & $-$59:55:08.53 & 17.306 & $-26\pm10$ & 29600 & $-$1.172 \\
A & 19 & 28947 & 19:11:04.724 & $-$59:54:56.61 & 16.832 & $-36\pm8$ & 23200 & $-$0.979\\
B & 1 & 4964 & 19:10:49.893 & $-$60:04:08.04 & 14.552 & $-47\pm8$ & 10700 & $-$0.258 \\
B & 2 & 49317 & 19:11:15.879 & $-$60:06:00.17 & 13.889 & $-35\pm8$ & 8600 & $+$0.028 \\
B & 3 & 5455 & 19:10:52.734 & $-$60:03:36.84 & 17.371 & $-34\pm6$ & 28700 & $-$1.147 \\
B & 4 & 5487 & 19:10:57.554 & $-$60:03:35.02 & 16.766 & $-43\pm10$ & 21800 & $-$0.935 \\
B & 5 & 5134 & 19:11:09.624 & $-$60:03:56.23 & 15.615 & $-40\pm8$ & 16200 & $-$0.706 \\
B & 6 & 4672 & 19:11:21.548 & $-$60:04:31.46 & 17.181 & $-33\pm8$ & 27900 & $-$1.123 \\
B & 7 & 5201 & 19:11:26.748 & $-$60:03:51.85 & 17.615 & $-37\pm7$ & 29900 & $-$1.181 \\
B & 8 & 5865 & 19:11:21.577 & $-$60:03:14.90 & 17.195 & $-44\pm9$ & 25100 & $-$1.037 \\
B & 9 & 7843 & 19:11:14.678 & $-$60:021:55.92 & 15.317 & $-41\pm8$ & 16000 & $-$0.693 \\
B & 10 & 6284 & 19:11:29.780 & $-$60:02:54.90 & 17.155 & $-36\pm6$ & 28600 & $-$1.142 \\
B & 11 & 10257 & 19:11:07.698 & $-$60:00:53.00 & 14.075 & $-37\pm10$ & 9300 & $-$0.075 \\
B & 12 & 10625 & 19:11:21.006 & $-$60:00:45.37 & 17.701 & $-24\pm9$ & 30700 & $-$1.201 \\
B & 13 & 8672 & 19:11:32.600 & $-$60:01:30.83 & 17.790 & $-33\pm9$ & 33500 & $-$1.266 \\
B & 14 & 10711 & 19:11:28.425 & $-$60:00:43.55 & 17.418 & $-31\pm7$ & 30200 & $-$1.189 \\
C & 1 & 11609 & 19:10:36.021 & $-$60:00:28.10 & 15.311 & $-37\pm9$ & 14600 & $-$0.613 \\
C & 2 & 14664 & 19:10:41.367 & $-$59:59:39.90 & 13.972 & $-7\pm9$ & 8300 & $+$0.086 \\
C & 3 & 14727 & 19:10:39.624 & $-$59:59:39.13 & 14.253 & $-33\pm9$ & 9600 & $-$0.119 \\
C & 4 & 35186 & 19:10:13.293 & $-$60:00:03.25 & 14.443 & $-34\pm9$ & 10600 & $-$0.239 \\
C & 5 & 35662 & 19:10:22.119 & $-$59:59:30.30 & 15.010 & $-40\pm9$ & 14000 & $-$0.566 \\
C & 6 & 35499 & 19:10:10.920 & $-$59:59:41.57 & 14.895 & $-35\pm9$ & 13400 & $-$0.518 \\
C & 7 & 36242 & 19:10:23.020 & $-$59:58:47.17 & 15.024 & $-31\pm7$ & 14100 & $-$0.570 \\
C & 8 & 36480 & 19:10:23.013 & $-$59:58:30.37 & 16.899 & $-35\pm8$ & 24800 & $-$1.028 \\
C & 9 & 36502 & 19:10:18.896 & $-$59:58:28.74 & 14.863 & $-35\pm9$ & 13200 & $-$0.505 \\
C & 10 & 36830 & 19:10:02.476 & $-$59:58:03.50 & 17.388 & $-33\pm8$ & 30700 & $-$1.202 \\
C & 11 & 38095 & 19:10:26.616 & $-$59:56:23.29 & 15.358 & $-35\pm9$ & 16100 & $-$0.700 \\
C & 12 & 38087 & 19:10:18.120 & $-$59:56:23.89 & 17.190 & $-37\pm8$ & 30500 & $-$1.197 \\
C & 13 & 32470 & 19:10:22.172 & $-$59:55:54.36 & 14.681 & $-27\pm8$ & 10800 & $-$0.271 \\
C & 14 & 28695 & 19:10:35.134 & $-$59:55:08.70 & 14.329 & $-$ & 9900 & $-$0.155 \\
C & 15 & 38504 & 19:10:15.308 & $-$59:55:41.62 & 14.823 & $-39\pm9$ & 12900 & $-$0.155 \\
C & 16 & 39008 & 19:10:25.034 & $-$59:54:41.75 & 17.178 & $-15\pm11$ &  31400 & $-$1.219 \\
C & 17 & 38889 & 19:10:09.836 & $-$59:54:56.10 & 14.897 & $-48\pm9$ & 13400 & $-$0.520 \\
C & 18 & 38963 & 19:10:03.450 & $-$59:54:46.56 & 14.264 & $-17\pm10$ & 9600 & $-$0.110 \\
\hline
\end{tabular}
\end{center}
\end{table*}
\clearpage

The list of targets, their coordinates, $V$ magnitudes, and
the relevant parameters extracted from the color-magnitude diagram and
our spectra are given in Table \ref{targetlist}.
The approximate effective temperatures were obtained from the $U-V$
colors by \citet{Moman02}, by applying the color-temperature relation
derived from the temperatures spectroscopically measured by \citet{mosw00}
on a sample of HB stars in the same cluster
as in Fig. \ref{teffM00}.
Combining the dispersion of the points around the fitted line in
Fig. \ref{teffM00}, and the photometric errors in $U-V$ we estimated
the error in effective temperature to be approximately 5\%.

During each night, up to two pairs of 1800s exposures were secured
in each field with grism 1400V+18, as shown in Table
\ref{frames}. The slit width was 0\farcs5
and the resulting resolution 1.2~\AA.
The exposures were always collected in pairs and subsequently summed
(see \S \ref{capRVmeasures}) except on the 3rd night for field A,
when only one 1800s exposure was acquired.
Just before each pair
of exposures, a slit image (without grism)
was taken, and these frames were used in the correction of
instrumental effects that affected the data (\S \ref{capcorr}).
The bias, flat, and lamp images were acquired before and at the end
of each observing night.
The observed spectral range was $\approx 1300$~\AA\ wide but, due to
the different positions of the slits within the masks, the
  central wavelengths of the spectra varied.
The bluer spectra reached $\approx 4200$~\AA\ in
the blue edge encompassing the $\mathrm{H_{\gamma}}$
line and the 4471~\AA\ HeI doublet,
while others extended on the red side up to $\approx
6000$ \AA .  The $\mathrm{H_{\beta}}$ line was always inside
the observed spectral range, except for star 28695.
This cool target was excluded from our analysis due to the lack of strong
lines in the spectral range.

\begin{figure}
\begin{center}
\resizebox{\hsize}{!}{\includegraphics{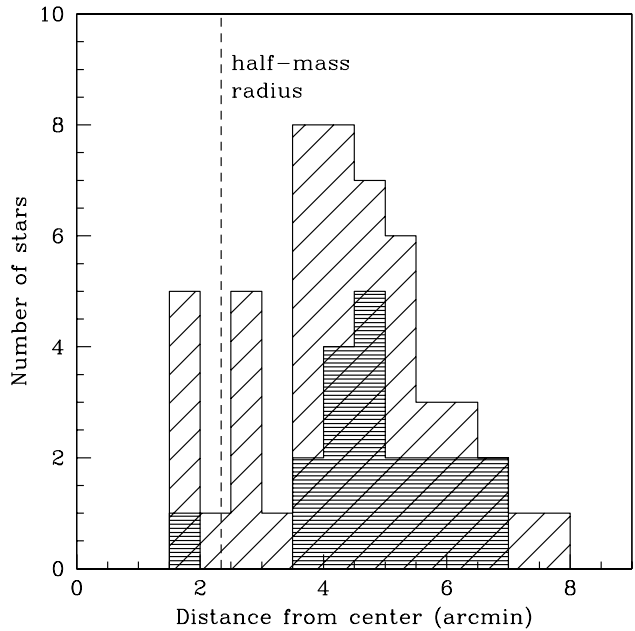}}
\caption{Radial distribution of the observed stars. The dark shaded
area indicates hot stars (T$_{\mathrm{eff}}\geq20000$~K). The
half-mass radius from Harris (1996) is also indicated.}
\label{radist}
\end{center}
\end{figure}

Data reduction from the multi-spectrum frames
to the one-dimensional calibrated spectra was
performed with standard MIDAS\footnote{ ESO-MIDAS is the acronym for
 the European Southern Observatory Munich Image Data Analysis System
 that is developed and maintained by the  European Southern
 Observatory (http://www.eso.org/projects/esomidas/)}
 procedures. All slitlets
were extracted from the full frames (bias, wavelength calibration,
 flat field, science spectra) and reduced independently.
The wavelength calibration (wlc) was performed with the HeNeHgCd lamp
images, fitting a $3^\mathrm{rd}$ order polynomial to the dispersion relation.
We re-binned the 2D frames to constant wavelength steps of 0.25~\AA /pix.
The extraction region for target and sky varied strongly from star to star because
different parameters affected the selection, mainly the crowding
level
and the object magnitude.
We tried to extract all the spectra with an optimal extraction routine
\citep{Horne86},
but for some stars (mainly the brightest ones) the procedure gave bad results,
so in these cases we opted for a simple average over the
extraction region.

\begin{table}
\begin{center}
\caption{List of $1800$s exposures acquired each night. The UT of the start
of the exposures (hour and minutes) is indicated,
and also the averaged seeing (variations between the exposures in the same night
are within 0\farcs1).}
\label{frames}
\begin{tabular}{c| c c c c}
\hline
\hline
field & \multicolumn{4}{|c}{night} \\
\hline
 & 12 & 13 & 14 & 15 \\
\hline
  & & 2:44 & & \\
A & 8:43 & 3:15 & 9:34 & 7:47 \\
  & 9:14 & 3:56 & 8:18 & \\
  & & 4:27 & & \\
\hline
  & 5:55 & & & \\
B & 6:27 & 5:08 & 8:33 & 8:58 \\
  & 6:59 & 5:39 & 8:59 & 9:29 \\
  & 7:30 & & & \\
\hline
  &     & 6:39 & & \\
C & -- & 8:04 & 7:05 & 6:36 \\
  & -- & 8:34 & 7:33 & 7:06 \\
  &     & & 7:59 & \\
\hline
Average seeing & 0\farcs9 & 1\farcs2 & 1\farcs3 & 1\farcs4 \\
\hline
\end{tabular}
\end{center}
\end{table}

No flux calibration was applied to the spectra,
because they were continuum-normalized
for the measurement of radial-velocity variations.


\begin{figure}
\begin{center}
\resizebox{\hsize}{!}{\includegraphics{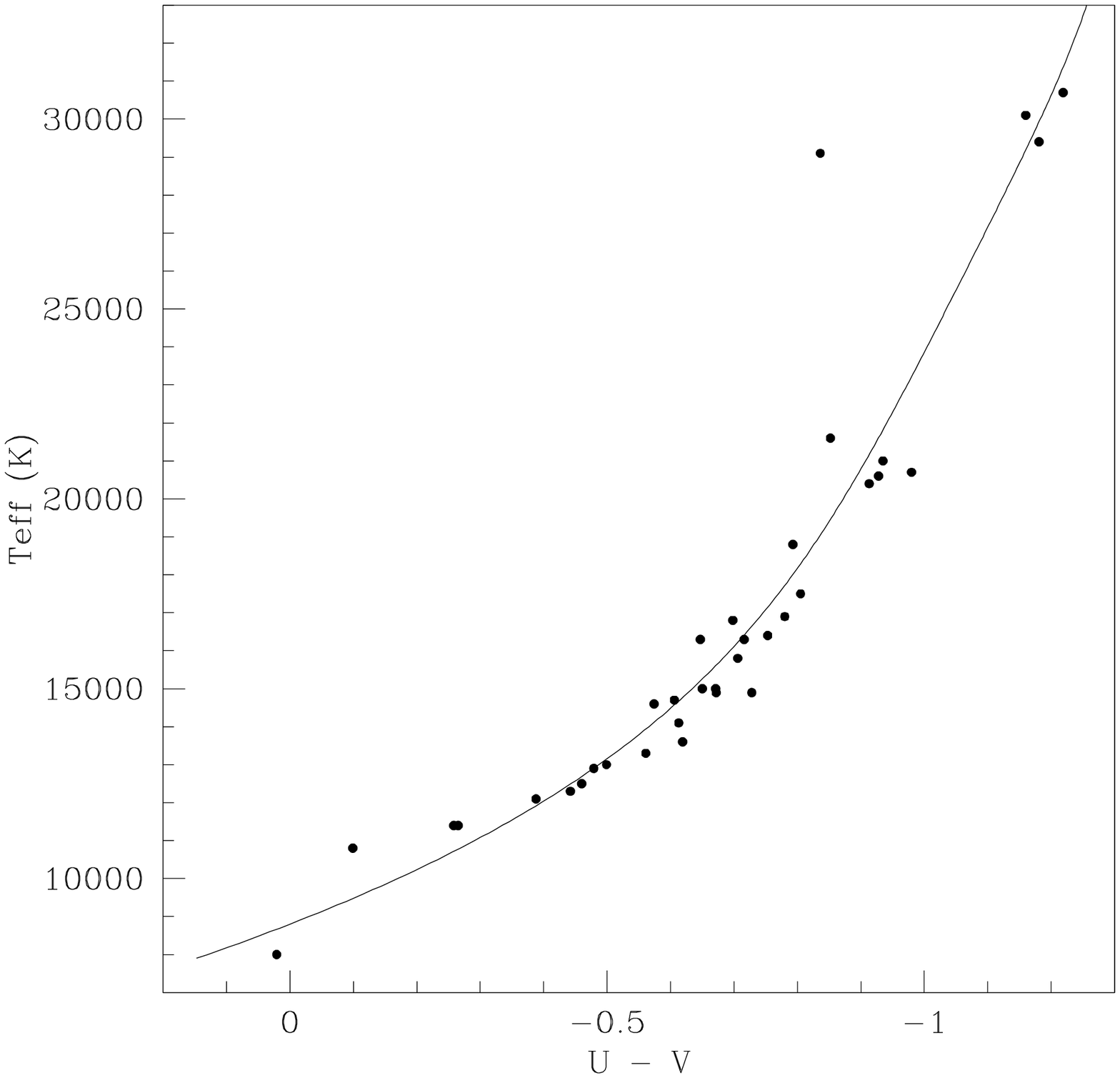}}
\caption{Plot used to derive the approximate effective temperatures of our targets.
The points are the effective temperatures measured by \citet{mosw00} using
models with solar metallicity, plotted against
the $U-V$ color in the \citet{Moman02} photometry. Seven stars listed
in \citet{mosw00} have been excluded since their identification in the
\citet{Moman02} photometry is dubious.
The line indicates the T$_{\mathrm{eff}}$ vs. color relation obtained
by a fit of the data points, used to estimate the temperatures of the stars observed in
our work. The point at $U-V\approx$-0.8,~T$_{\mathrm{eff}}$~$\approx$~29000 K were excluded
from the fit.}
\label{teffM00}
\end{center}
\end{figure}

\section{Measures}
\label{capmeasures}

\subsection{RV-variation measurement}
\label{capRVmeasures}
Radial velocities (RVs) were measured with the cross-correlation
(CC) technique \citep{TonEDavis79}, using the {\it fxcor}
IRAF\footnote{IRAF is distributed by the National Optical Astronomy
Observatories, which are operated by the Association of Universities
for Research in Astronomy, Inc., under cooperative agreement with the
National Science Foundation.} task.
Before co-adding each pair of spectra we measured the RV
variation between them, in order to verify that no significant
variation had occurred and that no information was lost due to
averaging. The
summed spectra will be referred to with the field and the night of
observation (from 12 to 15), with an additional letter (a or b) when
two pairs were acquired in the same night and field.  The A14 spectra
come from a single 1800s exposure, since only one frame was
acquired that night, and in our analysis we took their
lower S/N into account.

\begin{figure*}
\begin{center}
\includegraphics[width=17cm]{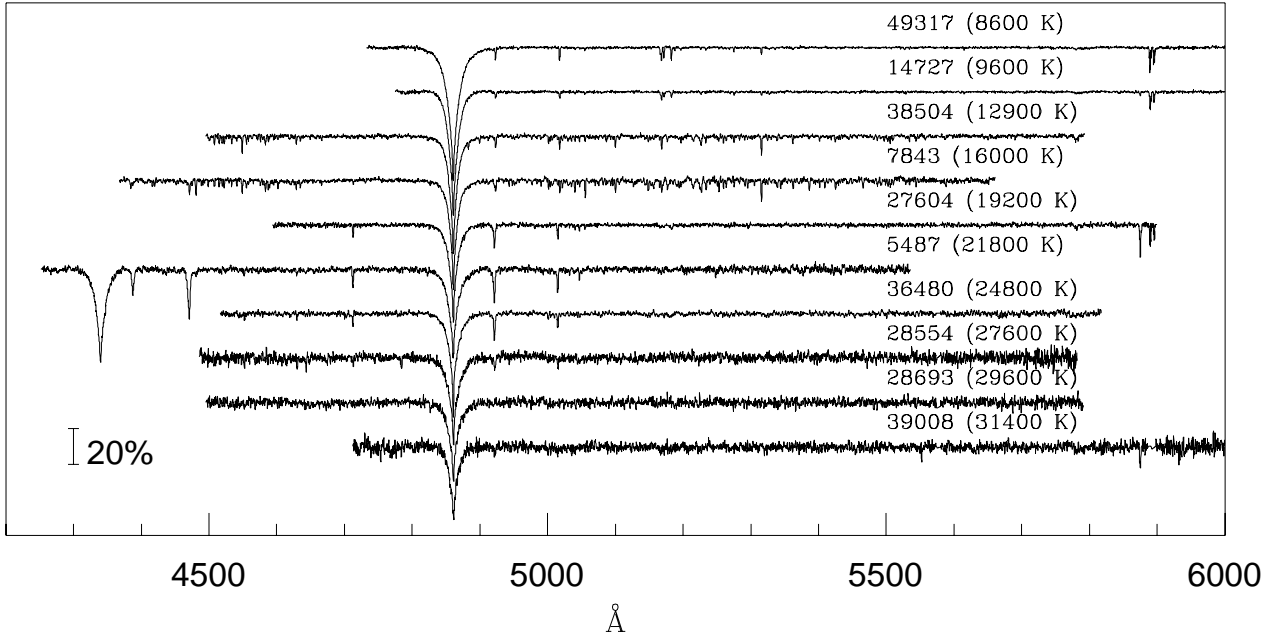}
\caption{A sample of the collected spectra,
normalized and ordered by increasing $\mathrm{T_{eff}}$.
The increasing noise with temperature is clearly visible, as well as
the large quantity of metallic lines for stars with $11\,500 \leq
\mathrm{T_{eff}} \leq 18\,000$.
}
\label{spectrafig}
\end{center}
\end{figure*}

For each target star all the spectra were cross-correlated,
thus performing 10 CCs for each star in fields A and B and
6 in field C, covering different temporal intervals from one hour to
$3.1$ days. In each measurement the first spectrum (in temporal
order) was assumed as template (the choice of which of the two
spectra is used as template affects only the sign of the RV variation).

For each RV measurement the cross-correlation function (CCF) was
computed and the position of its peak determined with a
Gaussian fit \citep[see for example][ for a description of the
procedure]{Recio04,Dubat90}.  In the measurement concerning Balmer lines,
the CCFs have Gaussian shapes only in the peak but not in the wings,
and sometimes they are asymmetric; whereas the CCFs obtained from weak
lines are approximated well by a Gaussian, but they are
often distorted due to low S/N. Therefore the fit
was restricted to the central portion around the peak, but
different fits were tried, varying the width of the CCF region fitted,
in order to find the best estimate for the RV
variations.

Our analysis focused on the $\mathrm{H_{\beta}}$ line, and 432
measurements were performed in the $4\,830-4\,890$ \AA\ spectral
range ($\mathrm{H_{\beta}}$ line with full wings).
We also compared
these results with measurements from other parts of the spectrum,
both to clarify the influence of noise on the measured RV variations
and to overcome the intrinsic weakness of the CC technique for blended lines in
double spectra \citep[see for example][]{Zucker94}.
Therefore, although our survey
was intended to be based on analysis of the $\mathrm{H_{\beta}}$,
all the measurements were repeated, cross-correlating the entire
spectra (with and without $\mathrm{H_{\beta}}$) and
$\mathrm{H_{\gamma}}$ (alone and with $\mathrm{H_{\beta}}$) when it
fell inside the spectral range, for a total number of $1\,532$
CCs. The presence of many metallic lines in the spectra of stars with
$11\,500 \leq \mathrm{\mathrm{T_{eff}}} \leq 18\,000$ K (see
Fig.~\ref{spectrafig}), due to
radiative levitation of heavy elements \citep{Glasp89,Behr03},
gave good CCFs even without the H lines.
On the
other hand, the low S/N and the lack of useful lines in the spectra of
hot stars (T$_{\rm eff} \ge 18,000$K)
 usually prevented cross-correlating the entire spectra
without $\mathrm{H_{\beta}}$, and then only spectral sections with
the strongest helium lines were cross-correlated in place of
these measurements, with quite uncertain results (see \S
\ref{ccerror}). Fourier filters of various shapes \citep{BraEWhite71}
were applied to almost all the noisy spectra, obtaining better
CCFs but unchanged results for the most part.

\begin{figure*}
\begin{center}
\includegraphics[width=17cm]{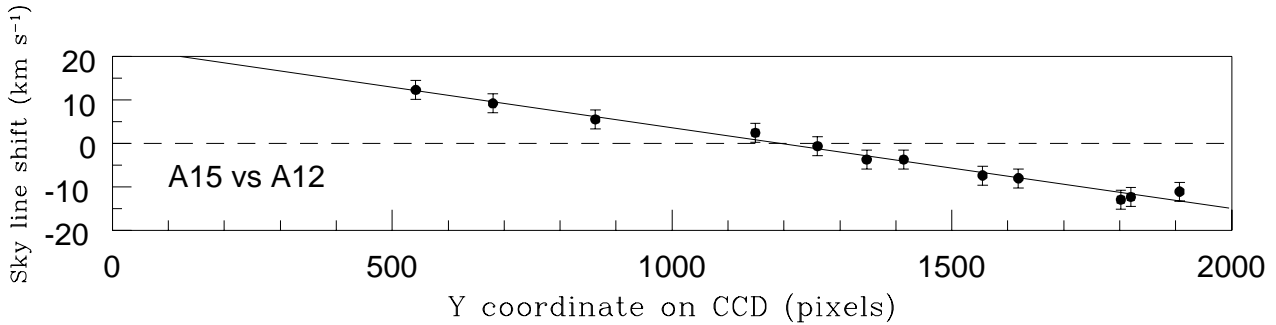}
\caption{An example of the corrections to re-position the 5577 \AA\
sky line to its laboratory wavelength. The differences (in km
s$^{-1}$) in the position of the sky line in the spectra A15 and A12
are plotted as a function of the Y-positions of the slits on the CCD
(in pixels).  The corrections for the few stars without the sky line
in the observed spectral range hwere obtained from their
Y-position by the least-square solution (straight line in the
figure). The error bar in the plot is the $1\sigma$ estimated for this
corrections (see Sect. \ref{err1step}).}
\label{5577plot}
\end{center}
\end{figure*}

The [OI] $5577$ \AA\ sky line was used as zero-point in order
to correct possible spectral shifts due to differences between lamp and
star spectra. The sky line fell outside the spectral range for 13
stars, preventing us from correcting the spectra before the
measurements.  We found that the differences in the sky line position in
the spectra between each pair of frames is a linear function of
Y-position (perpendicular to the dispersion direction) of the slits in
the masks (Fig. \ref{5577plot}). From the measured sky line
positions, we then calculated these differences for each pair of spectra,
and applied them as corrections to the RV variations measured with the
CCs. For the 13 stars missing the [OI] line, we calculated the
corrections from their Y-coordinates by means of the least-square
fit obtained by using the spectra with the [OI] line.


\subsection{Correction of the RV measures}
\label{capcorr}

\begin{figure}
\begin{center}
\resizebox{\hsize}{!}{\includegraphics{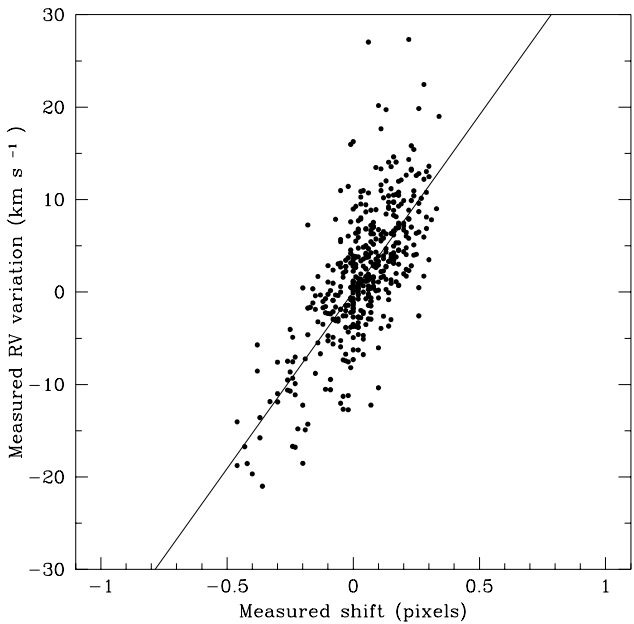}}
\caption{Displacements of stars with respect to the center of the slit
plotted against the RV variations measured from CCs. The straight line
indicates the instrumental relation 1 pixel = 38.2 km s$^{-1}$
given by the dispersion of the spectra.}
\label{biaII}
\end{center}
\end{figure}

After the skyline-based correction mentioned above,
we were forced to correct the measured RV variations again, since they
showed a clear correlation with the displacement (along the dispersion
direction) of the stars with respect to the center of the slit.
The effect resulted in an evident systematic error, up to 10--12
  km s$^{-1}$,
of the same order of magnitude for all the stars within the
same pair of frames (usually with a dependence on Y-position on CCD,
probably due to rotation of the mask and/or the field).
In the slit frames we measured the position of the stars
relative to the center of the slits with a Gaussian fit of the stellar
profile parallel to the dispersion direction.
Then, we obtained the differences in these positions between
pairs of frames, and translated them from pixels to km~s$^{-1}$
with the instrumental relation 38.2 km s$^{-1}$ pixel$^{-1}$,
in order to evaluate the effect of this shift on the RV measurements.
Actually, it is not the real stellar profile that has been fitted,
but the star profile convolved with the narrow slit,
that generates the spectrum on the CCD.
In fact, for narrow slits, the relation between the movement of the star behind them
and the induced RV variation should be flatter than the dispersion relation,
but a shift of the center of the profile that
reaches the grism through the slit
is expected to induce an RV variation given by the dispersion relation,
with good approximation. This is confirmed by the plot in Fig.
\ref{biaII}, where we compared the two quantities (shift of the
profile and RV variation) for each star and each pair of frames: the
points show a good agreement with the expected relation indicated by
the straight line.
Although the plot confirms that it is a good solution for relating the
two quantities, it also possibly indicates that it is just a
first-order approximation, because a deviation from the straight line
for higher displacements can be seen. The real solution could be an
S-shaped function. Nevertheless we adopted a linear approximation, that
still gave good enough corrections for our purposes and our overall
errors, in order to correct the RV variations with quantities that
depend only on the measured displacements and not on the RV
variations themselves (as would be the case if, for example, we fit
the points in Fig. \ref{biaII} to obtain a higher-order relation).

\begin{figure}
\begin{center}
\resizebox{\hsize}{!}{\includegraphics{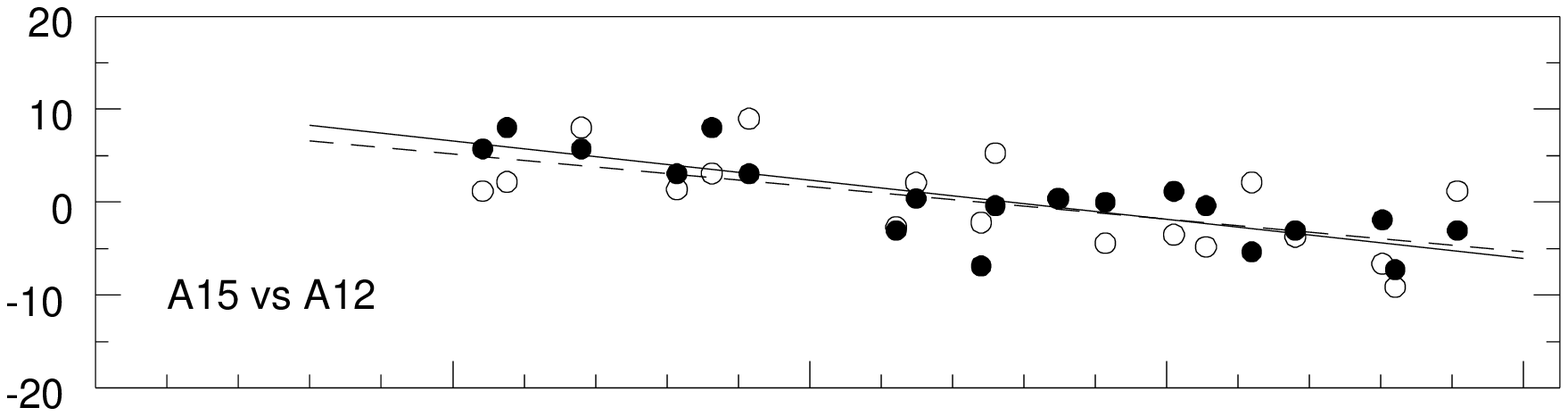}}
\resizebox{\hsize}{!}{\includegraphics{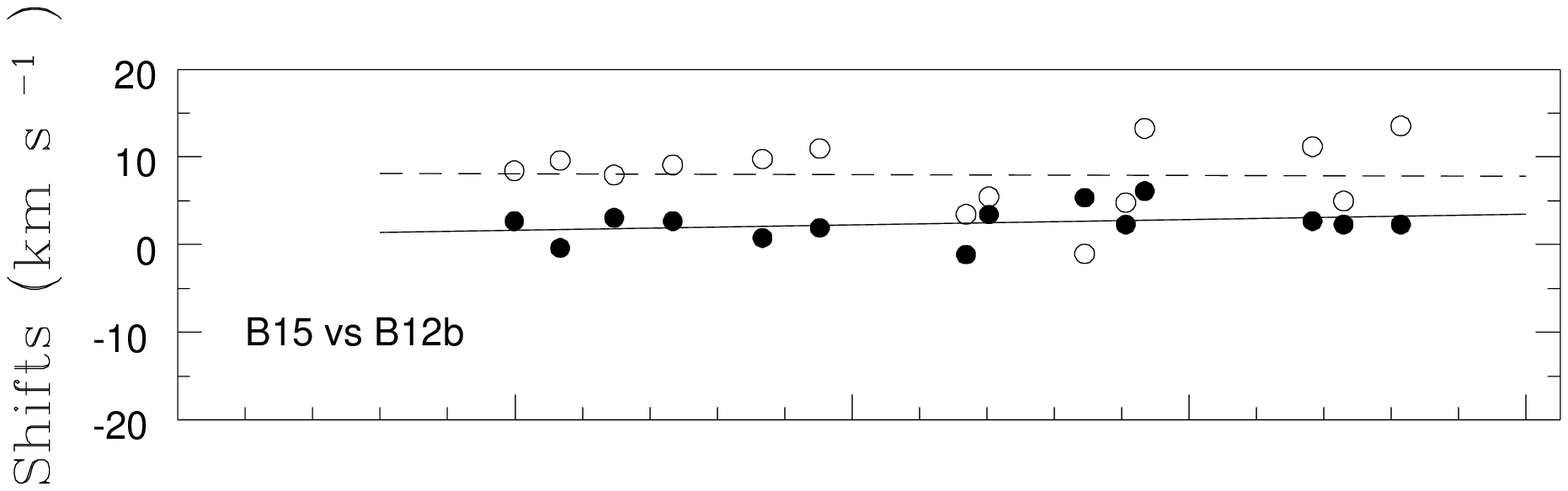}}
\resizebox{\hsize}{!}{\includegraphics{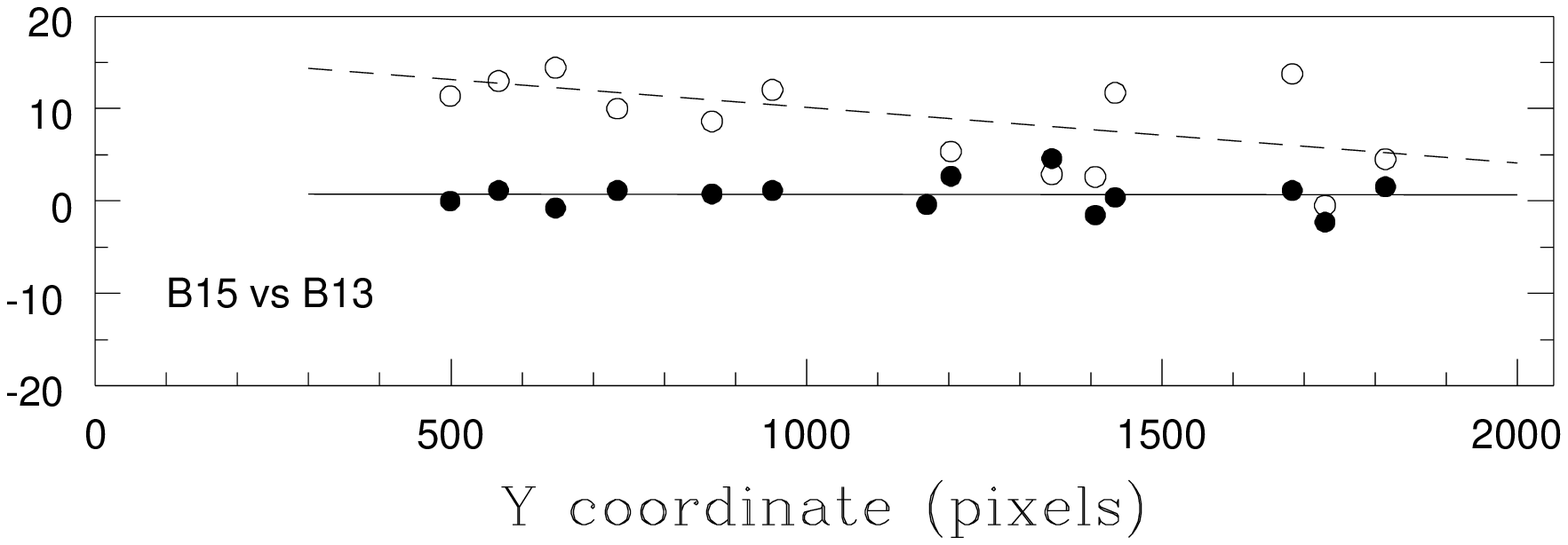}}
\caption{Comparison between the measured displacement of stars inside the slit
(solid points), translated from pixels to km s$^{-1}$, and the RV
observed variations (open points).
The lines are the least-square solutions for the two sets of data.
{\it Upper panel}: a typical plot (frames A15 and A12), in which the two sets
of data are very similar.
{\it Central Panel}: same plot involving frame B12b. A shift between the
two lines
is evident.
{\it Bottom panel}: same plot involving frame B13. The two sets of data
do not agree.}
\label{plot2ndstep}
\end{center}
\end{figure}

We then compared the displacements of the profiles and the RV variations
for each star, as a function of the Y coordinate on CCD. A typical
example is shown in Fig. \ref{plot2ndstep} (upper panel).
A dependence on slits'
Y-position can also be seen, but with a certain scatter due to the
random errors that affected the stellar fitting procedure in the
narrow slits. We preferred to derive the final corrections by applying
the value of the least-square fit
of these plots to each star
instead of the real shifts measured star-by-star,
in order to avoid introducing additional noise into the data.

For frames B12b and A13a, the slit image was not acquired
immediately before the exposures, but one hour before for frame B12b
and one hour after for A13a. In both cases, the corrections do not
agree with the RV variations observed, but a constant shift
between them is clear (Fig. \ref{plot2ndstep}, middle panel) and has of the
same amount for both frames. This indicates a time-dependent
instrumental movement (probably a shift of the mask in
its housing) that shifted the spectra slightly ($\approx 7\
\mathrm{km\ s}^{-1} = 0.2\ \mathrm{pixels}$).
Either way, the slit image for frames B12a and B12b was the same
(collected just before the first, B12a), and they were
taken in sequence, without any telescope-pointing change, and
calibrated with the same lamp image. For all these reasons,
it is reasonable to assume that
the observed shift of the sky line on the CCD between the frames
B12a and B12b was only due to
this time-dependent instrumental movement. Therefore we think that
we have been able to measure and correct this effect properly.

The corrections derived for frame B13 do not overlap the RV observed variations
(Fig. \ref{plot2ndstep}, lower panel).  Since it is the only
frame out of 14 for which this happens, its slit frame seems not to be reliable
for some reason. Probably the mask moved inside
its frame between the slit image acquisition and the spectra exposure.
We estimated a mask shift of 0.13--0.34 pixels with a rotation 0.00928
degrees. This is not unlikely and similar effects have been noted by
other observers before. Since we could not both find an explanation
and prove it, we preferred to exclude this frame from further
analysis.  The uncorrected RV variations observed are still useful and
give some information (see \S \ref{capresults}), although only in a
qualitative way.


\subsection{Absolute RV measures}
\label{capabsRV}

We measured absolute RVs in order to check the cluster membership of
the observed stars by means of CCs with the template star
\object{HD188112}, a binary sdB star with known ephemeris
\citep{Heber03}. The results are shown in Col. 7 of Table
\ref{targetlist}. The final absolute RVs have been corrected for sky
line position, as already described in \S \ref{capcorr}.
All the stars
show an absolute RV in agreement with that of the cluster
\citep[$-27.9$ km s$^{-1}$,][]{Harris96} within 2$\sigma$, and can
thus be
considered RV cluster members.


\section{Error analysis}
\label{caperror}

\begin{figure*}
\begin{center}
\includegraphics[width=17cm]{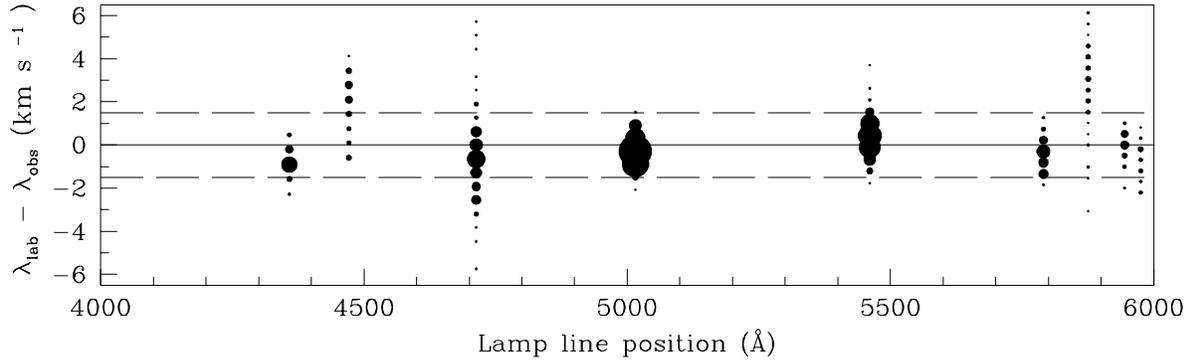}
\caption{Shift of lamp lines compared to their laboratory wavelength, in km s$^{-1}$.
Each circle is proportional to the frequency of the datum. The dashed lines
indicate the 1$\sigma$ dispersion.}
\label{calerror}
\end{center}
\end{figure*}

We performed an accurate error analysis in order to estimate the significance
of the measured RV variations. We analyzed all the main error sources,
and estimated a standard error $\sigma$ for each of them. The final errors
associated to the results presented in Table \ref{tabresults}
were obtained as their quadratic sum.

\subsection{Cross-correlation errors}
\label{ccerror}

The error in the CC procedure is obtained by the CC theory as
$$
\sigma_\mathrm{CC} = \frac{N}{8B (1+r)} \\
$$
where
$$
r = \frac{h}{\sqrt{2}\sigma_\mathrm{a}}
$$
and N is the number of bins, B the highest wavelength in which the CCF Fourier
transform
has an appreciable value, h is the height of the CCF peak, and $\sigma_\mathrm{a}$
the rms of the antisymmetric part of the CCF \citep[see][ for details]{TonEDavis79}.
This error is computed directly by the {\it fxcor}
IRAF task during each CC procedure. In the measurements with Balmer lines
it is usually between 0.5 km s$^{-1}$ (for high S/N spectra) and 2 km s$^{-1}$
(for asymmetric CCFs), but occasionally CCFs with bad profiles lead to
a CC error up to 5 km s$^{-1}$. The measurements without Balmer lines are affected
by much higher errors for hot stars (usually 3-5 km s$^{-1}$),
whereas for cooler stars the accuracy remained quite unchanged due to the
numerous metallic lines in their spectra.

\subsection{Wavelength calibration errors}
\label{wlcerror}

The precision and reliability of the wlc procedure, which is
  essential for our analysis, was
tested in each of the 186 lamp images used. These images were
calibrated with the coefficients obtained in the wlc procedure, and
the position of 9 bright lamp lines were measured with a Gaussian fit
and compared to their laboratory wavelength. This analysis
pointed out the level of inaccuracy in the wlc procedures that could
affect the RV measures.
The obtained data are plotted in Fig. \ref{calerror}.

No systematic error was found. Only the lines 4471 and 5875 \AA\ seem
calibrated
redder than the theoretical wavelength, but every night the mean shift of the
former was not greater than the wlc error, and the latter line lies in the
extreme red part of the spectra, never used in the CCs,
and between two
well-calibrated lamp lines. The dispersion of the data is in the range
1.3-1.7 km s$^{-1}$, slightly but not significantly varying among the
nights and
the three fields. They were considered as an estimate of the wlc error
$\sigma_\mathrm{wlc}$,
which has been counted twice (in quadrature) in the final error estimate
because in each CC two spectra are involved.

\subsection{Extraction and fit errors}
\label{fiterror}

Two additional sources of error are present in the measurement
procedure. The first one is related to the extraction of spectra from
the 2D images, as the choice of the aperture width can change
the profile of spectral lines (mostly in noisy spectra or in the presence
of cosmic spikes) slightly, and the CC procedure is sensitive enough to reveal
the differences.  The second, more important, is related to the
choice of the number of CCF points used to compute its Gaussian
fit. For distorted CCFs, this error is much greater than
$\sigma_\mathrm{CC}$, since the final result could depend on the
chosen fit.

These error sources were evaluated together by re-extracting all
the spectra in different manners and performing new measures with
different fits in the $\mathrm{H_{\beta}}$ wavelength.  We repeated
75\%\ of the measurements (326 out of 432), chosen in a stochastic way
in order to avoid selection effects in this choice.  The distribution
of the differences among these new measurements and the ones used for
this paper give the combined extraction and fit error
$\sigma_\mathrm{ex}$ (Table \ref{tabfiterror}).  We divided the
spectra in groups with similar S/N, i.e. hot and cool stars
($\mathrm{\mathrm{T_{eff}}} \leogr 20\,000$~K) and very hot stars in
field B ($\mathrm{T_{eff}} \approx 30\,000$~K). In field A, the
extraction and fit errors for the CCs involving the not-summed A14
spectra (much noisier) were evaluated separately and, as
expected, they came out much greater than the others.

\begin{table}
\begin{center}
\caption{
Standard deviations (km s$^{-1}$) of the differences between the
RV measurements obtained with different fits and spectral extractions
In field B the bluest stars have been distinguished from the
others, and ''single spectrum'' indicates the measurements involving the not-summed
spectra A14.}
\label{tabfiterror}
\begin{tabular}{c c c c}
\hline
$\sigma_\mathrm{ex}$ & \multicolumn{3}{c}{field} \\
\hline
 & A & B & C \\
\hline \hline
cool stars & 1.49 & 0.73 & 1.18 \\
hot stars &  2.48 & 2.11 & 2.69 \\
very hot stars &-- & 2.66 &-- \\
single-spectr. cool stars & 2.10 &-- &-- \\
single-spectr. hot stars & 3.46 &-- &-- \\
\hline
\end{tabular}
\end{center}
\end{table}

\begin{figure*}
\begin{center}
\includegraphics[width=17cm]{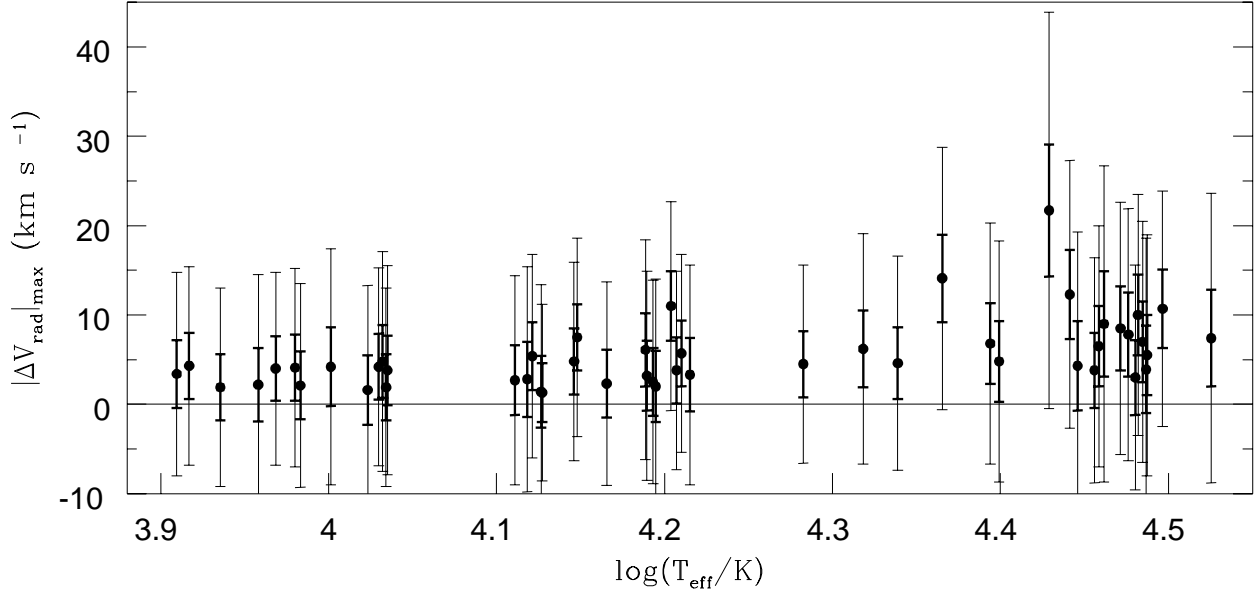}
\caption{Maximum RV variation for each star (in absolute value)
measured in H$_{\beta}$ wavelength, plotted as a
function of $\mathrm{\mathrm{T_{eff}}}$.
The thick errorbars are the $1\sigma$ errors, while the thinner bars
indicate the $3\sigma$.
There is no RV variation exceeding the $3\sigma$ error bars.}.
\label{plotHb}
\end{center}
\end{figure*}

\subsection{Sky line position errors}
\label{err1step}

The correction procedure described in \S \ref{capRVmeasures} introduced an
additional error, due to the uncertainty in the measurement of the
$5577$ \AA\ sky line
position. We assumed this error to be equal to the dispersion of the
corrections
around the least-square solutions when plotted against the slit Y-position, as
in Fig. \ref{5577plot}. We estimated $\sigma_\mathrm{sky} = 0.025$ \AA,
as expected for a fit error $\sigma = 0.1$ pixels. A similar value was
obtained by analyzing the Gaussian fits of the sky lines.
Then for the measurements in $\mathrm{H_{\beta}}$ we have
$$
\sigma_\mathrm{sky} = 0.025\ {\rm \AA} = 1.54\ \mathrm{km\ s}^{-1}
$$
to be considered twice (with proper error propagation),
as each correction involves two independent measurements.

\subsection{Correction to RV measures}
\label{err2step}

The corrections to the RV measurements described in \S \ref{capcorr}
have been an additional source of error. For each pair of frames,
we obtained the corrections by plotting the differences in stellar
positions as in Fig. \ref{plot2ndstep}. The standard deviations of
residuals with respect to the least-square solution were calculated for
each pair, and used as the estimated error $\sigma_\mathrm{corr}$
introduced by this procedure. It changed only slightly from one pair
to the other, being always $0.7 \leq\ \sigma_\mathrm{corr} \leq\ 1.4$
km s$^{-1}$.


\section{Results}
\label{capresults}

\begin{figure*}
\begin{center}
\includegraphics[width=17cm]{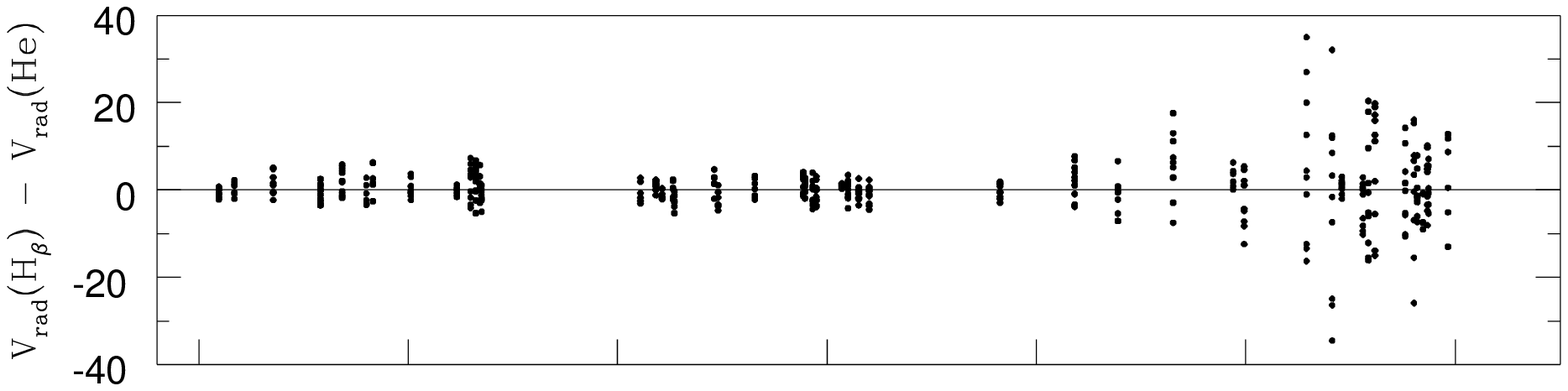}
\includegraphics[width=17cm]{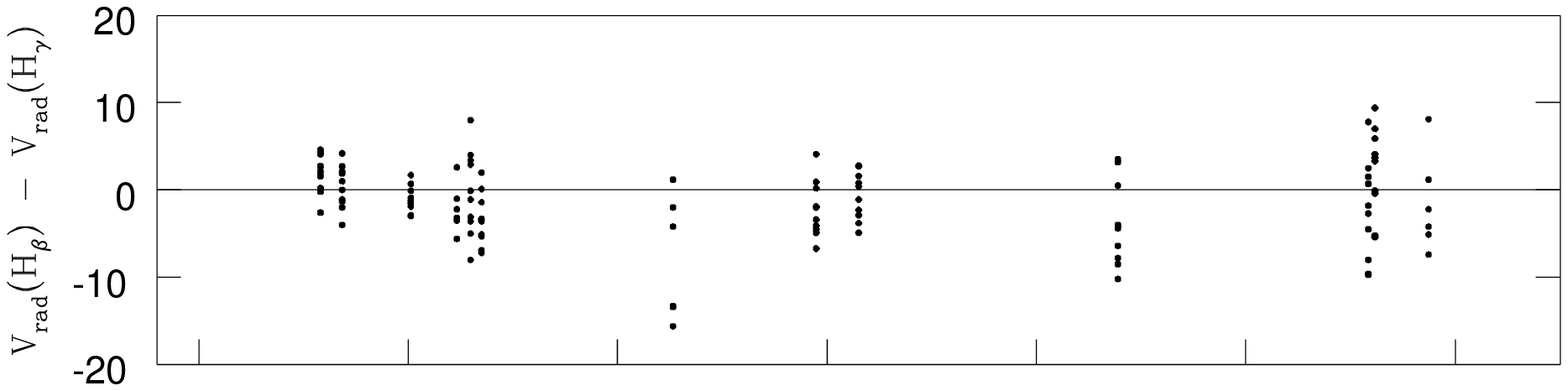}
\includegraphics[width=17cm]{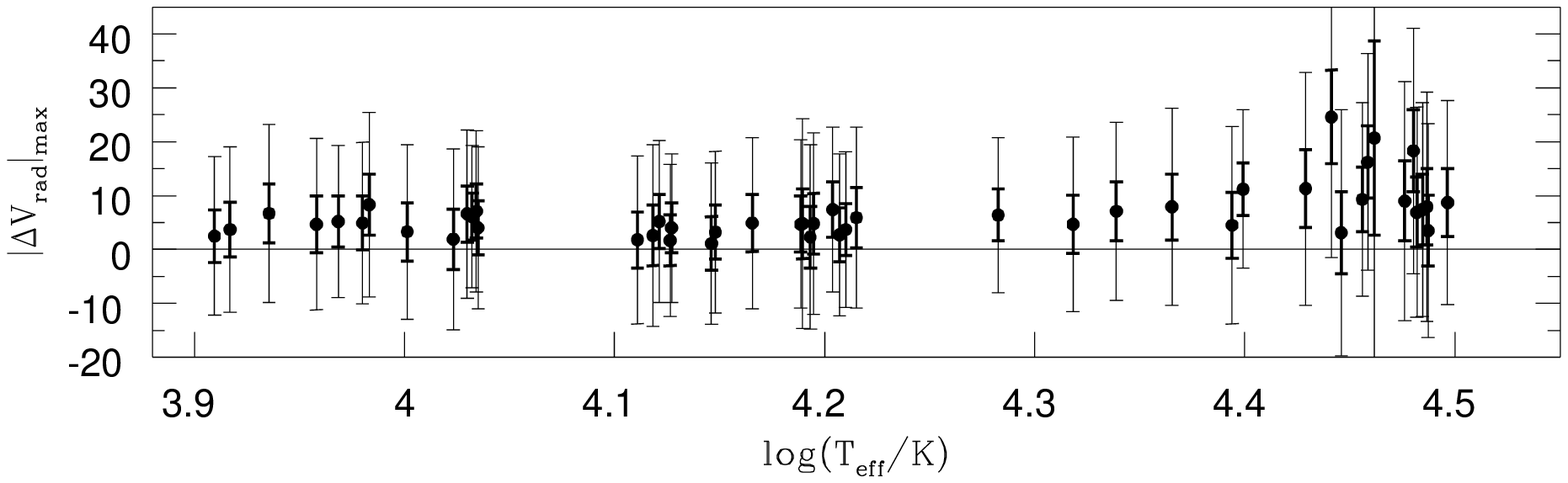}
\caption{Summary of the results in other wavelengths, plotted vs. effective
temperature of the stars. {\it Top panel}: differences between the results
in $\mathrm{H_{\beta}}$ and the weak lines. {\it Central panel}:
differences between the results
in $\mathrm{H_{\beta}}$ and $\mathrm{H_{\gamma}}$.
{\it Bottom panel}: maxima RV variations measured with weak lines.
The thick error bar indicates the 1$\sigma$ interval, and the thin
one is the $3\sigma$.}
\label{plotHe}
\end{center}
\end{figure*}

\subsection{RV variations with $\mathrm{H_{\beta}}$}

We obtained the most accurate and reliable RVs using the
$\mathrm{H_{\beta}}$ line.  The RV variations from all the possible
combinations of the observed spectra are listed in Table
\ref{tabresults}. For each star, Fig.  \ref{plotHb} shows the absolute
value of the measured maximum RV variation.

All the variations are small, usually below 10 km s$^{-1}$, and never
further than $3\sigma$ from zero; therefore {\it no RV variation can be
considered significant}.  The variations are only slightly larger in
the blue part of the HB, but with larger errors due to decreasing S/N
in the spectra.

The highest variation plotted in Fig. \ref{plotHb} (21.7 km s$^{-1}$) refers
to star 28231. We consider this datum interesting but particularly
dubious, since it was measured in frame A14 where we measured higher
variations, due to higher noise of the spectra, and star 28231
showed particularly distorted line profiles. This variation is not confirmed by the
measurements in other wavelengths, and this star shows no other remarkable
variation on the other nights. Therefore, it is highly probable that
the relatively high RV variation is only due to
noise-distorted $\mathrm{H_{\beta}}$ wing profile.

Table \ref{tabresults} also gives the results concerning the frame B13
(uncorrected for the star position inside the slit, see \S
\ref{capcorr}), although these have been excluded from our analysis
and from the plot in Fig. \ref{plotHb}. It can be seen that this
exclusion represents no substantial loss of information, since the
RVs appear scattered around a certain systematic value
different from zero, with no larger RV variations than
the typical $3\sigma$ interval.  Then even when including the data from the
uncorrected frame B13, we can confirm our general results.

\subsection{RV variations in other wavelengths}

The results with $\mathrm{H_{\gamma}}$, when available in the observed
spectral range, always confirm the ones using
$\mathrm{H_{\beta}}$. Occasionally the difference reaches $\approx 10$
km s$^{-1}$ (Fig. \ref{plotHe}, central panel), but this agrees
with a Gaussian distribution of the differences with a dispersion equal to
our estimated errors.

The measurements with weak metallic lines gave very good results
for cool stars (Fig. \ref{plotHe}, upper and bottom panel), because
of the large number of lines and the good S/N of the spectra. The RV
variations are always small, and the general trend confirms the
results we had using $\mathrm{H_{\beta}}$ (the differences are on
average below 3 km s$^{-1}$).  For hot stars the results obtained using weak lines
(mainly He lines) are not very reliable for many
reasons: they are not uniform, since the number of useful lines
changes from spectrum to spectrum; the wavelength is not clearly
determined, so the applied corrections introduce some
uncertainty, due to the translations from \AA\ to velocity
quantities; the lines are few and weak, and then easily distorted by
noise spikes.
Still all the RV variations are within the $3\sigma$ error bars,
although the errors are larger than in
the previously discussed measurements.
For some stars the maximum RV variations occasionally reach 30 km
s$^{-1}$ (Fig. \ref{plotHe}, bottom panel). It is worth noting that
these differences
always tend toward limiting high variations in
$\mathrm{H_{\beta}}$, never toward emphasizing them, and this
is also clear from the fact that, in spite of these great differences,
the maximum RV variations are on absolute value that is the same order of
magnitude as in $\mathrm{H_{\beta}}$ or just slightly higher. For
example, both the stars 28231 and 28947, which showed
the highest RV variations in $\mathrm{H_{\beta}}$, show no great variation in
 the weak line measurements.  Therefore the measurements with weak lines
confirm the results with $\mathrm{H_{\beta}}$, and indicate
that the higher RV variations observed with $\mathrm{H_{\beta}}$ are
simply due to random errors.

\subsection{Binary detection probability in our observations}
\label{capprob}

In order to better understand the significance of our results
we calculated the probability $d$ of detecting a binary in our observations
as a function of the period P. In order to relate the period and the
maximum semiamplitude of the RV curve, we assumed a circular orbit
and a mass of 0.5 M$_{\sun}$ for both components. These assumptions
are representative of the typical binary systems observed in
field sdB stars.

For each value of P we considered 50 possible values of $v\sin(i)$
(where $i$ is the inclination of the orbit
along the line of sight), and 50 possible values of the phase T$_{0}$
(equally distributed at constant step in the range $0 \leq \mathrm{T}_{0} \leq
\mathrm{P}$
and $0 \leq \mathrm{v\sin(i)} \leq 1$).
Then, we calculated how many of these 2500 binary systems would have been
detected in our observations, defining
 as ``detection'' an RV variation of more than
20 km s$^{-1}$ (our 3$\sigma$ for hot stars) between any two of our
observed epochs.
Finally we weighted the probabilities for the
three fields by the number of hot
stars observed in each, in order to derive the average detection
probability $d$ of our observations.

\begin{figure}
\begin{center}
\resizebox{\hsize}{!}{\includegraphics{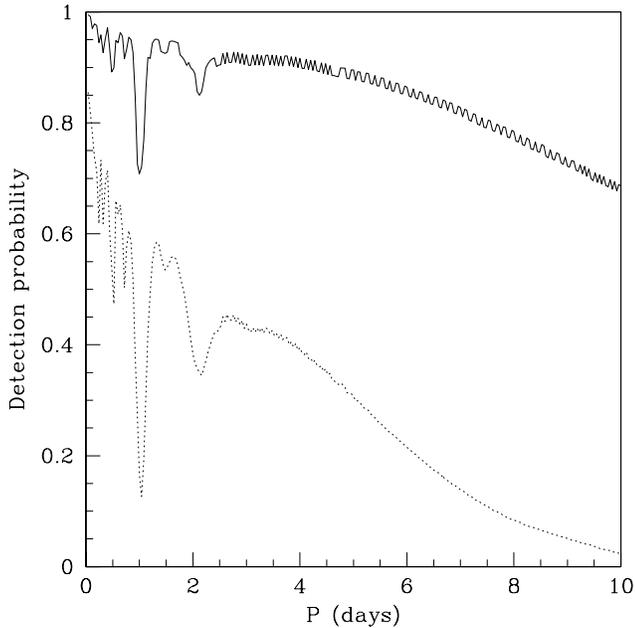}}
\caption{Probability of detecting a binary system in our observations
as a function of its period P. The solid line indicates the probability
calculated for an sdB star and a companion of 0.5 M$_{\sun}$, the dotted line
for a companion of 0.1 M$_{\sun}$, assuming
an EHB star of 0.5 M$_{\sun}$ in both cases.}
\label{probfig}
\end{center}
\end{figure}

The results are shown in Fig. \ref{probfig}.
The probability of detecting a binary
with periods P$<5$ days
is usually higher than 80\%, and
reaches 90\% for periods shorter than one day.
There is an evident loss
of sensitivity around P = 1 day, since this is the typical temporal
interval between two observing epochs.

We performed similar calculations also assuming a companion of
0.1 M$_{\sun}$. This kind of system is a minority among the binary
population in the field, but they do exist, as for example the
well-known \object{HW Vir} system \citep{Mezier86}.
As shown in Fig~\ref{probfig}, the detection probability for such
systems is very low in our survey, with the exception of the shortest (1-2 days)
period binaries.

\subsection{Binary fraction estimate}
\label{capfrac}
If the fraction of binaries in the sample is $f$
the probability of detecting N$_{\mathrm B}$ binaries out of a sample of N stars is:
$$
\mathrm{P} = \frac{\mathrm{N!}}{(\mathrm{N-N_{\mathrm{B}}})!\mathrm{N_{\mathrm B}}!}(df)^\mathrm{N_{\mathrm B}}(1-\mathrm{\bar{d}f})^{\mathrm{N-N_{\mathrm B}}}
$$
where \={d} is the probability of detection weighted for the period distribution of binaries.
In our survey N$_{\mathrm B}$=0 and N=18,
therfore,
$$
\mathrm{P} = (1-\mathrm{\bar{d}f})^{18} .
$$
Since we found no binaries, the probability P has a unitary maximum in
$f$=0.  Then, for increasing values of $f$, it falls rapidly to
zero. The exact shape of the function depends on the assumed period
distribution, which is very uncertain.  We performed the
calculations for two limiting cases, a flat distribution and a truncated
Gaussian centered on P=1 day and with log(P/days)=1 as FWHM, as also
assumed by \citet{Maxted01} and \citet{Napiw04}. This
distribution seems to follow the observations of field sdB stars,
although the known data are still too few and this is just a first
guess.  The results on P obtained with the two distributions are
extremely similar to within 1\%, because the decline of the Gaussian wings
in the second case compensates for higher (lower) values of the
sensitivity for short (long) periods, and the weighted mean is very
similar.  The value of P reaches 0.05 for $f$=0.20, in both
cases. Then we can conclude that the
binary fraction among EHB stars in \object{NGC6752} is lower then
20\% at a confidence level of 95\% .


\section{Discussion and Conclusions}
\label{discussion}

The results of Fig.  \ref{plotHb} clearly indicate that there is no
close binary system in our sample. The RV variations observed during
the four nights, with different temporal samplings, are all within the
estimated errors. These errors ($\approx 4-6$ km s$^{-1}$) are
significantly smaller than the
radial velocity variations we were expecting based on the results for
field sdB stars.

In the compilation by \citet{Moral03}, for all the 38 field sdBs with
known orbital periods, the RV semi-amplitude is always greater than 30
km s$^{-1}$. These binaries should have been easily detected in our
survey
at a 5-6 sigma level.
If the binary fraction of the EHB stars in
\object{NGC6752} is the same as among the field sdB stars listed in
\citet{Maxted01}, we would expect that $13\pm2$ of the 18 stars with
$T_e>$22,000~K in our sample should be close binaries; given the
period distribution of \citet{Moral03}, 80\% of them
should have a period $P<5$ days and, therefore, be easily detected with
our temporal sampling.
Indeed,  as shown in the previous section, our survey should be able to
detect 85\% of close binaries with $P<5$ days of the kind found among
the field sdB stars. I.e., if the binary fraction among the hot
cluster EHB stars were the same as among field sdB stars, we
would have expected to find 9 binaries with RV variations $>20$
km s$^{-1}$.

We found no significant RV variation at 3 sigma
level (15-18 km s$^{-1}$). Only 18 measurements showed a variation
exceeding 2 sigmas (10-12 km s$^{-1}$). However, assuming a Gaussian
distribution for our errors, we would expect that 19.6 measures among
our 432 RV difference estimates should exceed the 2 sigma level;
therefore, these variations too cannot be considered significant.  As
demonstrated in Sect. \ref{capfrac}, these results imply that the fraction of
close binaries among EHB stars in NGC6752 is $<20$\%, at a 95\%
confidence level.  This result is in sharp contrast with that of
\citet{Peterson02}, who concluded, from the measurement of radial
velocity variations of 30 EHB stars in \object{NGC6752}, that the
majority of them are close binaries.  \citet{Peterson02} unfortunately
have not published more details about their investigation so far, and
therefore, it remains to be established whether the different
conclusions are due to the lower S/N of \citet{Peterson02} data (who
used the multifiber Hydra spectrograph mounted on a 4m size
telescope) or to some physical reasons. Interestingly enough,
apparently, our sample of stars is different from their sample. In
particular, according to Peterson (priv. comm.), their EHB
stars are on average located in a less crowded, more external environment than
ours (though our stars are evenly distributed from $\sim$ 100\arcsec\
to $\sim$ 500\arcsec\ from the cluster center, i.e. out to about 3.5
half-mass radii, see Fig. \ref{radist}). We have only five
stars in common, of which only two are EHB stars and for which also
\citet{Peterson02} do not find evidence of radial velocity variations
(Peterson, priv. comm.).

In summary, we are forced to conclude that most ($>80$\%) of the hot EHB stars
in \object{NGC6752} are not close binaries. This result poses a number
of problems. First of all, it implies that the mechanism for the
formation of EHB stars in GCs does not involve a large envelope mass
loss enhanced by tidal interaction within a close binary
at variance with what happens for the field sdBs. Or, at
least, this cannot be the dominant formation mechanism.  In a sense,
this result is not completely unexpected.
 The typical companions of field sdB stars are relatively massive
stars \citep[0.3-0.5 M$_{\sun}$, see] []{Maxted04}. If the same binary
systems are the source of the EHB stars in GCs, they should originally have
had a mass that is on average larger ($\geq$ 1M$_{\sun}$) than the mass of a
single star at the TO ($\sim 0.8$ M$_{\sun}$) and, therefore, should be more
concentrated than the upper MS/RGB stars because of mass
segregation. Despite a number of attempts, there is no compelling
evidence at the moment that EHB stars are, on average,
more concentrated than cooler HB stars in the same clusters, as we would expect
if they were members of a binary system with similar properties
to the binaries producing field sdBs, and
as we find, e.g., for the blue stragglers in {\it all}
the observed GCs \citep{Piotto04}. Instead, recent surveys
\citep[e.g. by][ in \object{NGC2808}, a prototype EHB
  cluster]{Bedin00} have shown
that no clear radial gradient of the EHB stars can be detected.
 This empirical evidence, coupled with the results of the present
  paper, implies that: either 1) the companions of the EHB stars have very
  low masses (so that the sum of the main sequence mass of the EHB stars
  and that of the companion do not differ much from the mass of
a typical MS star), or  2) cluster EHB stars all have periods much longer
than 5 days, or 3) cluster EHB stars are not binaries.

In any case, we are forced to conclude that in GCs there are
different formation mechanisms of EHB stars with respect to the main
formation channels of field sdBs.
It is possible that the dynamical evolution of GCs have removed the
primordial binaries able to produce sdB stars, at least in the inner
part of the cluster. This is an interesting possibility. Indeed,
\citet{Piotto04} have shown that the frequency of blue stragglers
(BSs) in the cores of GCs is significantly smaller than the frequency
of field BSs, and that there is an anticorrelation between the
frequency of BSs and the GC total mass. \citet{Davies04}
suggest that this is likely due to the fact that in more massive
clusters, where the encounter probability is higher, binary evolution
is accelerated by similar mechanisms to those proposed by
\citet{Davies98} to explain the formation of pulsars in GCs. In this
way, the frequency of BSs (from primordial binaries) could have been
much higher in the past, but now there are much fewer binaries to
produce them via the merger channel.  However, the GC environment can
favor different formation mechanisms for the EHB stars.  Again,
something similar happens for the BSs. \citet{Piotto04} show
that the luminosity functions of the BSs in the core of massive
clusters differ significantly from those in less massive ones, and the
former is compatible with BSs mainly formed via collision of stars
(more probable in more massive clusters).

Interestingly enough,
\citet{Recio05} show that more massive clusters also show more
extended HBs.  If the \citet{Peterson02} results are
confirmed by a more extended survey. i.e. if at least a fraction of
the EHB stars in the outskirts of the clusters are indeed close
binaries (at variance with what we find in the more internal regions),
we will have further evidence of the effect of the environment on
the formation of EHB (or sdB) stars. This would also explain the sharp
difference between the frequency of close binaries among cluster EHB
stars and field sdBs. The results of the present paper and of
\citet{Peterson02} might suggest that two different channels of
production of EHB stars are at work, at least in some GCs. Again,
it is tempting to recall that there is compelling evidence of two
channels of BS production in the same GC.  It has been shown that
the radial distribution of BSs in \object{M3} \citep{Ferraro97} and in
\object{47Tuc} \citep{Ferraro04} are clearly bimodal and that
the BS properties in the inner regions and in the outskirts of this
cluster are also different.  \citet{Mapelli04} show that the radial
distribution of BS in 47Tuc supports the hypothesis that the BSs
produced in the center are caused mainly by collisions, while the BS
in the cluster outskirts (for $r>30$ core radii, i.e. for distances
from the center greater than 4 half mass radii) must all result from
the merge of primordial binaries, which survived because of the
much slower dynamical evolution in the external part of the
cluster. A similar scenario could also be at work for EHB stars, with
EHB stars in the external part of the cluster coming from the close
binary-evolution channel suggested for the production of field sdB
stars.  It would be interesting to use the appropriate dynamical and
stellar evolution models to investigate the possibility that EHB stars
in GCs can be formed preferentially by stellar collisions in their
more central parts.  Surely, our results and those of
\citet{Peterson02} call for a much more extended observing campaign to
search for close binaries among EHB stars in this and other GCs.

Another interesting possibility that needs to be (observationally)
explored is that cluster EHB stars are produced in binaries with
significantly longer periods or with significantly less
massive companion than in field sdBs. As shown in Fig.\ref{probfig}, our survey
could find only binaries with relatively large-mass companions, and
as it only has a 4-day temporal coverage, it is totally insensitive to
those binaries with periods longer than 10 days. We note that the
\citet{Han03} models for the production of sdB stars predict a large
number of long period ($\sim 100$ days) binaries. A follow-up
observing campaign is needed to verify if there are binaries like
these among cluster EHB stars; in any case, long-period binaries seem
to be a minority among field sdBs, so that the question to why
field and cluster EHB stars should form through different mechanisms
remains unanswered.

Of course, it is equally possible that dynamical evolution and binaries
have nothing to do with the formation of EHB stars in GCs.
Unlike field sdBs, which can have
progenitors with rather different masses,
all the cluster EHB stars start their evolution from a well
defined TO mass ($\sim 0.8$~M$_{\sun}$, slightly depending on cluster
age and metallicity), because of the very small
(compatible with zero) age and metallicity dispersion among stars in a
given GC. There must be an extremely well-tuned
(unlikely?) mass-loss mechanism to produce a star with the very small
core and envelope mass dispersion of the cluster EHB stars.
Other explanations are possible. As discussed in the introduction, recent
results on very massive clusters like \object{$\omega$ Centauri}
\citep{Piotto05} and \object{NGC2808} \citep{dantona05} seem to
indicate that the EHBs in GCs could represent the evolved population
of a second generation of stars formed by material enriched in He
because of the pollution by the ejecta of SNe and/or intermediate mass
AGBs from a first generation of stars.  In this hypothesis, it is
possible that more massive clusters are better able to keep part of
the ejecta, explaining the correlation found by \citet{Recio05}.  One
should keep in mind, however, that not only the total mass, but also
the concentration of a globular cluster determine the depth of its
gravitational potential and thereby its ability to retain enriched
material.

There is an interesting point to add before concluding this discussion.
As noted in the introductory section, \citet{Napiw04} found a much
lower (42\%) binary frequency among field sdBs than did
\citet{Maxted01}. On one hand, even with this low binary frequency, we
would expect to find 5 binaries among our sampled 18 EHB stars in
\object{NGC6752}. \citet{Napiw04} suggest that the difference
in mean apparent magnitude between their sample and the one of
\citet{Maxted01} may imply a difference in the populations sampled
by the two surveys. Although this has not yet been proven, attributing
the observed difference in binary fraction to a population difference
would mean that binaries are substantially less frequent among thick
disk-halo sdBs, and it would imply a trend with age and/or
metallicity.  On average, our cluster EHB stars are expected to be
older, and possibly more metal poor than the Napiwotzki et al. sample,
and, apparently, the fraction of binaries among them is even smaller
than among the Napiwotzki et al. sample.  It is currently unclear if
these differences are related to the different environments, ages, or
metallicities, or to a combination of these parameters.  Supporting
evidence for the possible influence of abundance differences comes
from the observed abundance anomalies in GC red giants \citep[][ and
references therein]{Catelan05, Gratton04}, which are not seen in
field red giants \citep{Gratton00}. As suggested by many
in the literature \citep[e.g.][]{VandESmith88, Sweigart97, Dantona02},
these abundance anomalies (either primordial, or due to mixing
effects) can affect the RGB evolution and, consequently, have an impact
on the properties of HB stars including temperatures, luminosities,
gravities, and pulsation characteristics.


\begin{acknowledgements}
GP acknowledges the support by MIUR within the PRIN2003 program.
RAM acknowledges support from the Chilean Centro de Astrof\'{\i}sica FONDAP
(No. 15010003). We thank Ron Webbink for useful discussions and for
pointing out the problem of the small core and envelope mass 
dispersion of EHB stars when trying to interpret their origin in terms
of mass loss within a binary system.
ARB acknowledges the support of the European Space Agency.
We thank the staff at Paranal observatory for their support
during the observations.
\end{acknowledgements}


\bibliographystyle{aa}

\bibliography{biblio}

\begin{table*}
\begin{center}
\caption{RV variations for target stars.}
\label{tabresults}
\begin{tabular}{c | r r r r r r r r r r}
\hline
star & 13a vs 12 & 13b vs 12 & 13b vs 13a & 14 vs 12 & 14 vs 13a & 14 vs 13b & 15 vs 12 & 15 vs 13a & 15 vs 13b & 15 vs 14\\
\hline \hline
1a & $-5.5\pm4.7$ & $-10.0\pm4.5$ & $-2.0\pm4.2$ & $-3.1\pm5.2$ & $-2.5\pm4.9$ & $-1.1\pm5.0$ & $-4.0\pm5.5$ & $0.0\pm4.3$ & $3.3\pm4.9$ & $3.6\pm5.0$ \\
2a & $-3.2\pm4.2$ & $-3.3\pm4.0$ & $-0.3\pm3.9$ & $-4.2\pm4.4$ & $-0.9\pm3.9$ & $-2.4\pm4.2$ & $-3.0\pm4.0$ & $0.4\pm3.9$ & $0.6\pm3.7$ & $1.0\pm3.9$ \\
3a & $0.0\pm4.2$ & $1.5\pm4.1$ & $2.2\pm3.8$ & $-1.6\pm4.4$ & $-1.1\pm3.9$ & $-2.4\pm4.2$ & $2.4\pm4.1$ & $2.5\pm3.8$ & $0.6\pm3.7$ & $0.7\pm3.9$ \\
4a & $-0.4\pm4.2$ & $-2.2\pm4.1$ & $-1.8\pm3.7$ & $-1.0\pm4.4$ & $-0.6\pm3.9$ & $-0.8\pm4.1$ & $-2.0\pm4.0$ & $-1.6\pm3.8$ & $0.0\pm3.7$ & $-1.0\pm3.9$ \\
5a & $1.9\pm4.3$ & $-1.4\pm4.3$ & $-3.2\pm3.9$ & $2.7\pm4.4$ & $0.9\pm3.9$ & $2.1\pm4.1$ & $-0.3\pm4.0$ & $-2.3\pm3.9$ & $1.2\pm3.7$ & $-3.1\pm3.9$ \\
6a & $-2.3\pm4.7$ & $-2.5\pm4.5$ & $0.4\pm4.3$ & $5.7\pm7.3$ & $9.0\pm5.9$ & $2.3\pm7.2$ & $4.7\pm4.6$ & $7.5\pm4.3$ & $6.6\pm4.6$ & $-2.2\pm7.6$ \\
7a & $0.2\pm4.2$ & $-1.7\pm4.1$ & $-2.2\pm3.8$ & $-0.4\pm4.4$ & $-0.7\pm4.0$ & $-0.5\pm4.0$ & $-3.2\pm4.0$ & $-3.4\pm3.8$ & $-1.2\pm3.7$ & $-2.7\pm3.8$ \\
8a & $2.8\pm4.2$ & $1.2\pm4.1$ & $-1.5\pm3.8$ & $2.1\pm4.4$ & $-0.6\pm3.9$ & $-1.1\pm4.2$ & $0.8\pm4.0$ & $-2.7\pm3.7$ & $-0.6\pm3.8$ & $-1.6\pm3.8$ \\
9a & $2.0\pm4.2$ & $-1.7\pm4.1$ & $0.0\pm3.7$ & $-1.5\pm4.4$ & $-0.3\pm3.9$ & $-1.8\pm4.1$ & $-2.0\pm4.0$ & $-0.1\pm3.7$ & $-0.4\pm3.9$ & $-0.1\pm3.8$ \\
10a & $2.7\pm4.8$ & $-2.3\pm4.5$ & $-5.1\pm4.3$ & $5.4\pm5.2$ & $0.5\pm5.3$ & $5.8\pm5.1$ & $4.1\pm4.5$ & $0.3\pm4.6$ & $6.2\pm4.3$ & $-1.5\pm4.7$ \\
11a & $0.6\pm4.4$ & $-2.2\pm4.0$ & $-2.9\pm3.9$ & $4.5\pm4.4$ & $4.0\pm4.0$ & $4.8\pm4.1$ & $0.8\pm4.1$ & $0.3\pm3.8$ & $3.2\pm3.8$ & $-3.8\pm3.9$ \\
12a & $-3.5\pm4.2$ & $-6.1\pm4.1$ & $-2.6\pm3.7$ & $2.0\pm4.4$ & $5.4\pm3.9$ & $5.9\pm4.1$ & $-2.6\pm4.0$ & $0.8\pm3.9$ & $3.5\pm3.8$ & $-4.4\pm3.9$ \\
13a & $2.1\pm4.2$ & $-1.6\pm4.1$ & $-3.8\pm3.9$ & $-1.0\pm4.5$ & $-3.0\pm4.1$ & $-1.0\pm4.2$ & $-1.2\pm4.1$ & $-3.4\pm3.9$ & $0.7\pm3.7$ & $-0.2\pm4.0$ \\
14a & $3.1\pm4.2$ & $1.9\pm4.0$ & $-1.5\pm3.7$ & $1.3\pm4.4$ & $-1.6\pm3.9$ & $-2.5\pm4.1$ & $-2.0\pm4.0$ & $-4.5\pm3.7$ & $-3.4\pm3.7$ & $-3.0\pm4.1$ \\
15a & $-2.0\pm4.9$ & $-9.8\pm4.7$ & $-6.0\pm4.3$ & $13.5\pm6.5$ & $18.6\pm6.3$ & $21.7\pm7.4$ & $3.9\pm4.5$ & $7.7\pm4.7$ & $12.3\pm4.3$ & $-10.8\pm7.5$ \\
16a & $-0.6\pm4.2$ & $-3.3\pm4.1$ & $-2.6\pm3.8$ & $-1.0\pm4.5$ & $-0.1\pm4.0$ & $0.5\pm4.1$ & $-0.3\pm4.0$ & $0.6\pm3.8$ & $3.1\pm3.8$ & $1.1\pm3.8$ \\
17a & $5.3\pm4.9$ & $-3.5\pm4.5$ & $-6.7\pm4.4$ & $-11.3\pm5.3$ & $-12.3\pm5.0$ & $-9.9\pm5.0$ & $-1.8\pm4.9$ & $-5.4\pm4.5$ & $0.3\pm5.0$ & $7.5\pm4.9$ \\
18a & $-0.6\pm5.8$ & $-7.3\pm4.8$ & $-8.5\pm4.7$ & $-4.6\pm5.2$ & $-4.9\pm5.9$ & $-3.4\pm5.5$ & $-3.6\pm5.5$ & $-7.4\pm4.3$ & $2.3\pm4.5$ & $5.5\pm5.1$ \\
19a & $-4.3\pm4.8$ & $-6.4\pm4.5$ & $-1.6\pm4.3$ & $-7.2\pm6.2$ & $-5.2\pm5.1$ & $-3.5\pm5.1$ & $5.1\pm4.6$ & $9.9\pm4.5$ & $11.1\pm4.2$ & $14.1\pm4.9$ \\
\hline \hline
star & 12b vs 12a & 13 vs 12a & 13 vs 12b & 14 vs 12a & 14 vs 12b & 14 vs 13 & 15 vs 12a & 15 vs 12b & 15 vs 13 & 15 vs 14 \\
\hline
1b & $3.5\pm2.4$ & $-2.2$ & $-12.2$ & $0.9\pm3.8$ & $-3.3\pm3.7$ & $15.8$ & $-0.5\pm3.9$ & $-4.2\pm3.7$ & $9.0$ & $-1.6\pm3.4$ \\
2b & $-0.8\pm2.4$ & $-2.6$ & $-7.3$ & $-1.6\pm3.7$ & $-0.5\pm3.7$ & $13.6$ & $0.4\pm3.8$ & $1.8\pm3.6$ & $10.3$ & $1.9\pm3.7$ \\
3b & $-1.7\pm3.4$ & $-6.0$ & $-10.4$ & $-6.5\pm4.5$ & $-4.9\pm4.3$ & $12.0$ & $-1.0\pm4.2$ & $0.2\pm4.3$ & $11.4$ & $5.6\pm4.3$ \\
4b & $-0.7\pm3.2$ & $0.0$ & $-5.1$ & $-4.4\pm4.3$ & $-4.2\pm4.4$ & $7.3$ & $-0.1\pm4.2$ & $0.0\pm4.2$ & $7.9$ & $4.6\pm4.0$ \\
5b & $1.2\pm2.4$ & $0.5$ & $-6.3$ & $-4.7\pm3.8$ & $-5.7\pm3.7$ & $7.0$ & $0.4\pm3.8$ & $-0.7\pm3.7$ & $6.8$ & $5.1\pm3.6$ \\
6b & $-1.8\pm3.5$ & $-0.8$ & $-4.6$ & $-2.7\pm4.4$ & $-1.8\pm4.8$ & $8.8$ & $1.3\pm4.5$ & $3.3\pm4.1$ & $9.5$ & $4.3\pm5.0$ \\
7b & $-7.3\pm3.7$ & $-12.7$ & $-11.2$ & $-7.8\pm4.7$ & $0.4\pm4.5$ & $14.0$ & $-4.8\pm4.4$ & $0.8\pm4.5$ & $16.0$ & $2.3\pm4.5$ \\
8b & $-1.4\pm3.2$ & $0.4$ & $-3.8$ & $3.8\pm4.7$ & $4.8\pm4.5$ & $14.3$ & $-3.2\pm4.3$ & $-1.8\pm4.2$ & $4.2$ & $-4.7\pm3.9$ \\
9b & $-2.7\pm2.5$ & $-3.3$ & $-6.3$ & $-11.0\pm3.9$ & $-8.6\pm3.7$ & $4.0$ & $-8.4\pm3.6$ & $-5.8\pm3.7$ & $2.3$ & $2.2\pm3.4$ \\
10b & $-2.1\pm3.2$ & $2.9$ & $0.2$ & $-1.5\pm4.6$ & $1.5\pm4.5$ & $6.8$ & $-3.8\pm4.2$ & $-0.9\pm4.4$ & $2.1$ & $-1.7\pm4.6$ \\
11b & $-1.2\pm2.4$ & $1.3$ & $-3.0$ & $1.5\pm3.9$ & $2.8\pm4.0$ & $11.6$ & $2.9\pm3.6$ & $4.0\pm3.6$ & $9.3$ & $1.4\pm3.8$ \\
12b & $-1.1\pm3.5$ & $0.7$ & $-4.7$ & $-0.5\pm4.6$ & $-0.4\pm4.5$ & $9.7$ & $1.1\pm4.9$ & $3.9\pm4.9$ & $10.9$ & $2.5\pm4.7$ \\
13b & $1.1\pm3.8$ & $3.6$ & $-2.6$ & $3.2\pm5.1$ & $1.3\pm5.3$ & $9.4$ & $-3.8\pm4.5$ & $-5.4\pm4.5$ & $-0.4$ & $-7.4\pm5.4$ \\
14b & $0.9\pm3.2$ & $6.8$ & $0.9$ & $-1.0\pm4.3$ & $-2.2\pm4.5$ & $3.3$ & $3.0\pm4.2$ & $2.1\pm4.1$ & $3.6$ & $2.8\pm4.1$ \\
\hline \hline
\end{tabular}
\begin{tabular}{c |  r r r r r r}
star & 14a vs 13 & 14b vs 13 & 14b vs 14a & 15 vs 13 & 15 vs 14a & 15 vs 14b \\
\hline
1c & $-0.3\pm3.7$ & $-2.3\pm3.8$ & $-2.0\pm3.7$ & $-0.9\pm3.8$ & $-0.6\pm4.0$ & $1.4\pm3.4$ \\
2c & $0.1\pm3.7$ & $-1.9\pm4.0$ & $-1.7\pm3.7$ & $-4.3\pm3.7$ & $-4.2\pm3.7$ & $-2.2\pm3.5$ \\
3c & $-1.3\pm3.8$ & $-2.0\pm3.9$ & $-0.4\pm3.6$ & $-2.1\pm3.8$ & $-0.8\pm3.8$ & $-0.1\pm3.5$ \\
4c & $1.4\pm3.8$ & $-0.2\pm4.0$ & $-1.5\pm3.6$ & $-0.1\pm3.8$ & $-1.6\pm3.9$ & $0.0\pm3.4$ \\
5c & $2.5\pm3.7$ & $4.8\pm3.7$ & $1.7\pm3.6$ & $3.2\pm4.0$ & $0.3\pm3.7$ & $-2.0\pm3.4$ \\
6c & $1.4\pm4.0$ & $0.4\pm4.2$ & $-0.3\pm3.6$ & $0.3\pm3.8$ & $-0.9\pm3.7$ & $-0.6\pm3.4$ \\
7c & $3.4\pm3.9$ & $-1.4\pm3.8$ & $-4.9\pm3.6$ & $-4.3\pm3.8$ & $-7.5\pm3.7$ & $-2.8\pm3.7$ \\
8c & $-1.9\pm4.5$ & $3.9\pm4.9$ & $5.9\pm4.5$ & $6.2\pm4.5$ & $6.8\pm4.5$ & $-0.8\pm4.5$ \\
9c & $5.4\pm3.8$ & $1.8\pm3.7$ & $-3.6\pm3.7$ & $3.2\pm3.7$ & $-2.1\pm3.7$ & $1.4\pm3.5$ \\
10c & $4.2\pm4.7$ & $0.5\pm5.0$ & $-2.3\pm4.5$ & $-3.9\pm4.5$ & $-5.5\pm4.5$ & $-4.5\pm4.2$ \\
11c & $-0.7\pm3.8$ & $0.4\pm3.7$ & $1.1\pm3.6$ & $3.1\pm3.8$ & $3.8\pm3.7$ & $2.9\pm3.3$ \\
12c & $-1.4\pm4.5$ & $-7.0\pm4.5$ & $-4.4\pm4.5$ & $-4.0\pm4.7$ & $-0.8\pm4.5$ & $3.2\pm4.5$ \\
13c & $-1.4\pm3.8$ & $0.4\pm3.7$ & $1.9\pm3.7$ & $0.6\pm3.7$ & $1.6\pm3.7$ & $0.3\pm3.3$ \\
15c & $2.7\pm3.9$ & $2.0\pm3.7$ & $-0.8\pm3.7$ & $1.1\pm3.7$ & $-1.6\pm4.0$ & $-0.9\pm3.4$ \\
16c & $10.7\pm4.4$ & $3.5\pm4.6$ & $-5.1\pm4.5$ & $5.0\pm4.5$ & $-3.9\pm4.4$ & $3.2\pm4.2$ \\
17c & $-0.2\pm3.7$ & $0.5\pm3.7$ & $0.6\pm3.6$ & $-1.3\pm3.8$ & $-0.9\pm3.7$ & $-1.6\pm3.3$ \\
18c & $2.6\pm3.7$ & $4.1\pm3.7$ & $1.5\pm3.6$ & $1.2\pm3.8$ & $-1.9\pm3.8$ & $-3.4\pm3.4$ \\
\hline
\end{tabular}
\end{center}
\end{table*}

\end{document}